\documentclass[a4 paper,11 pt]{article}
\pdfoutput=1 

\usepackage{jheppub} 

\usepackage[T1]{fontenc}
\usepackage{amsmath}
\usepackage{physics}
\usepackage{cancel}
\usepackage{caption}
\usepackage{subcaption,longtable,stmaryrd}
\usepackage{comment}

\usepackage{subcaption}
\usepackage{float}

\DeclareSymbolFont{usualmathcal}{OMS}{cmsy}{m}{n}
\DeclareSymbolFontAlphabet{\mathcal}{usualmathcal}

\usepackage{bigints}
\usepackage{multicol}
\usepackage{blindtext}
\usepackage{tikz}
\usepackage{enumitem}
\usepackage{amsmath, amssymb, graphics, setspace}
\usepackage{float}
\newcommand{\mathsym}[1]{{}}

\newcommand{\unicode}[1]{{}}
\usepackage[mathscr]{euscript}
\DeclareSymbolFont{rsfs}{U}{rsfs}{m}{n}
\DeclareSymbolFontAlphabet{\mathscrsfs}{rsfs}
\DeclareSymbolFont{rmlargesymbols}{OMX}{mdbch}{m}{n}
\DeclareMathSymbol{\rmintop}{\mathop}{rmlargesymbols}{82}
\DeclareMathSymbol{\rmointop}{\mathop}{rmlargesymbols}{72}
\usepackage[customcolors,shade]{hf-tikz}

\usepackage{graphicx}

\definecolor{darkbrown}{rgb}{0.787, 0.26, 0.187}
\definecolor{alizarin}{rgb}{0.82, 0.1, 0.26}
\usepackage{tikz}
\title{
Krylov Complexity for Open Quantum System: Dissipation and Decoherence}
\author[a]{Arpan Bhattacharyya,}
\author[b]{Sayed Gool,} 
\author[b,c]{S. Shajidul Haque}
\affiliation[a]{Department of Physics, Indian Institute of Technology, Gandhinagar,
Gujarat-382355, India}
\affiliation[b]{The Laboratory for Quantum Gravity \& Strings, Department of Mathematics and Applied Mathematics, University of Cape Town, Cape Town, South Africa}
\affiliation[c]{The National Institute for Theoretical and Computational Sciences, Private Bag X1, Matieland,
South Africa}

\emailAdd{ abhattacharyya@iitgn.ac.in, GLXMOH001@myuct.ac.za, shajidhaque@uct.ac.za}

\abstract{We investigate Krylov complexity in open quantum systems using Lindblad master equations for bosonic bath models, with particular emphasis on the Caldeira–Leggett model. Krylov complexity is computed from the moments of the two-point function within the standard master equation framework. For the damped harmonic oscillator, the results reveal clear dissipative features in Krylov complexity. In the Caldeira–Leggett model, in the high-temperature limit, we find that Krylov complexity saturates in the full system and reproduces the expected dissipative behavior when the decoherence term is suppressed in the master equation. Conversely, when the dissipative term is suppressed, the contribution from decoherence exhibits the familiar oscillatory dynamics of the coherent system, along with additional novel features. However, Krylov complexity appears insensitive to the onset of decoherence, as no clear distinctive signature is observed. We attribute this to the fact that Krylov complexity is defined in the Krylov basis, which does not coincide with the conventional basis typically used to study decoherence.} 

\begin{document}
\maketitle
\flushbottom

\section{Introduction}
\label{sec:intro}
Understanding the mechanisms of dissipation \cite{Ferialdi_2017,Ramazanoglu_2009} and decoherence \cite{Schlosshauer_2019,Habib:1998ai} in open quantum systems is a fundamental challenge in physics, with far-reaching implications for fields ranging from low-temperature phenomena and early-universe cosmology to modern applications like quantum computing \cite{horodecki2013fundamental, Skrzypczyk_2014,Lesgourgues:1996jc,PhysRevA.52.R2493,Aharonov_2000,Ashida:2020dkc}. Decoherence occurs when a quantum system interacts with its environment, leading to a loss of phase coherence. For a comprehensive review of this subject, interested readers are referred to \cite{Schlosshauer:2019ewh, Zurek:2003zz}. Dissipation, in contrast, involves the irreversible transfer of energy from the system to its environment, often resulting in the damping of quantum dynamics. A useful quantity for characterizing dissipation in a quantum system is the probability amplitude \cite{Xu:2018bhd}, while the degree of decoherence—or, equivalently, the deviation of an open quantum system from a pure state—can be quantified using the decoherence function \cite{Zurek:2003zz,open}.\\

In recent years, quantum circuit complexity \cite{Susskind:2014moa, Jefferson:2017sdb, Nielsen:2005mkt} has gained momentum as a powerful tool for exploring the properties and behavior of quantum systems, capturing the interest of researchers across multiple disciplines. The concept first rose to prominence in the context of the AdS/CFT correspondence and black hole physics, where it was developed as an extension of entanglement entropy \cite{Ryu:2006bv} to probe the interior of a black hole \cite{Susskind:2014moa,Susskind:2014rva,Brown:2015bva,Brown:2015lvg,Carmi:2016wjl}. However, its foundations were already well-established in quantum information theory, where it has served as a measure of the computational resources required to prepare or simulate quantum states \cite{Baiguera:2025dkc}. Today, circuit complexity is widely applied to diverse quantum systems \cite{Jefferson:2017sdb, Hackl:2018ptj, Khan:2018rzm, Chapman:2018hou, Ali:2018fcz, Ali:2018aon, Ali:2019zcj}, including condensed matter physics \cite{Liu_2020, Camargo_2019, Craps:2023rur}, cosmology 
\cite{Bhattacharyya:2020rpy, Bhattacharyya:2020kgu,Haque:2021kdm}, astrophysics \cite{Dixit:2023fke,Jha:2025tde} and quantum computing \cite{Baiguera:2025dkc}.  Interested readers are referred to these reviews \cite{Chapman:2021jbh, Bhattacharyya:2021cwf,Baiguera:2025dkc} for more details.\\

Recent studies have shown that circuit complexity is sensitive to decoherence \cite{Bhattacharyya:2021fii,Bhattacharyya_2023,Bhattacharyya_2024,Bhattacharyya:2025cxv,Haque:2021hyw}, a process that disrupts quantum coherence and affects the performance of quantum systems in various contexts \cite{Hornberger,Schlosshauer_2019,BRE02}. This sensitivity makes circuit complexity a valuable diagnostic tool for understanding how environmental interactions impact quantum systems. More recently, an alternative measure, known as Krylov complexity, has been proposed as a complementary approach to quantifying the complexity of quantum dynamics \cite{Parker:2018yvk, Barbon:2019wsy, Avdoshkin:2019trj, PhysRevB.102.085137, Balasubramanian:2022tpr}. Unlike circuit complexity, which counts the minimum number of quantum gates needed to construct a state, Krylov complexity focuses on the evolution of operators in the Krylov space \cite{Parker:2018yvk}, offering a unique perspective on the dynamical behavior of quantum systems.\\

Given the sensitivity of circuit complexity to decoherence, it is natural to explore whether Krylov complexity exhibits similar behavior, particularly in the presence of both decoherence and dissipation—two fundamental processes that degrade quantum information in open quantum systems. Decoherence arises from interactions with an external environment, causing a loss of quantum coherence, while dissipation involves energy exchange, often leading to the damping of quantum states. Determining whether Krylov complexity can serve as a robust probe of these phenomena is of great interest, as it could offer new insights into the behavior of complex quantum systems under realistic, non-ideal conditions.\\

We investigate the sensitivity of Krylov complexity to decoherence and dissipation, with the aim of assessing its usefulness as a reliable measure for studying these processes. Focusing on quantum systems such as the damped harmonic oscillator and the Caldeira–Leggett model, we explore the potential of Krylov complexity as a diagnostic tool. We further compare its performance with circuit complexity and evaluate its ability to capture the effects of environmental interactions. Our results have implications for the study of open quantum systems and contribute to the broader understanding of complexity in quantum mechanics.\\
Krylov complexity is defined in the Krylov subspace, originally developed to address large-scale linear algebra problems \cite{doi:10.1137/1036123}. It offers a novel approach to studying operator growth, with a key advantage being its ability to reduce high-dimensional systems to a one-dimensional chain problem. Krylov complexity has proven particularly useful in diagnosing quantum phase transitions.
In maximally chaotic systems, complexity grows at most exponentially \cite{Parker_2019,Barbon:2019wsy,Rabinovici_2022, Rabinovici_2021, Bhattacharyya:2023dhp}, whereas in open quantum systems it typically exhibits rapid initial growth followed by late-time saturation, with a maximum value lower than that of the corresponding closed system \cite{Bhattacharya_2023,Liu_2023,Bhattacharyya:2023grv}. For more applications of Krylov complexity, readers are referred to these reviews \cite{Nandy_2025,Rabinovici:2025otw} and to some recent works where it is studied for non-Hermitian systems \cite{Bhattacharya_2022,Bhattacharya_2023, Liu_2023, Bhattacharyya:2023grv,Bhattacharjee:2023uwx, NSSrivatsa:2023pby,Bhattacharya:2023yec, Bhattacharya:2024hto,Chakrabarti:2025hsb, Medina-Guerra:2025wxg, Medina-Guerra:2025rwa, Baggioli:2025knt}, which motivated the study presented in this paper.  \\

In this study, we apply Krylov complexity to investigate features of open quantum systems, focusing on systems exhibiting dissipative and decoherent effects, such as the damped harmonic oscillator and the Caldeira-Leggett model \cite{Caldeira:1982iu, Bhattacharyya_2023}. The Caldeira-Leggett model describes a system coupled to a thermal bath of non-interacting harmonic oscillators, with dynamics governed by a Markovian master equation. This equation captures the slow, systematic evolution of the system compared to the environment, a hallmark of quantum Brownian motion.\\

The paper is organized as follows. In Sec.~ (\ref{KryComp}), we outline the details of Krylov complexity and its computation from the moments of the two-point function. In Sec.~(\ref{DampedOsc}), we discuss the damped harmonic oscillator model and present its Krylov complexity results. In Sec.~(\ref{CLmod}), we introduce the Caldeira-Leggett master equation, its assumptions, and its features, and analyze the associated Krylov complexity and its implications. Finally, in Sec.~(\ref{Discussion}), we summarize our findings and discuss their broader implications for future research. In the appendices, we have provided some additional details.
\section{Krylov Complexity} \label{KryComp}
In this paper, we investigate the growth of Krylov complexity in open quantum systems. We begin with a brief review of Krylov complexity, followed by an overview of the moments method used to compute it \cite{Parker_2019, dymarsky2020quantum, Dymarsky_2021,vish123}. Krylov complexity is particularly valuable for its ability to project high-dimensional problems onto a lower-dimensional Krylov subspace. Our analysis starts in the Heisenberg picture of operator evolution, where time evolution naturally introduces the Liouvillian superoperator and leads to the construction of the Krylov basis. To understand this framework, consider the Heisenberg equation in the operator picture:
\begin{equation}
    \partial_t O(t) = i[H, O(t)], \quad O(t) = e^{-iHt} O(0) e^{iHt}.
\end{equation}

Using the Baker-Campbell-Hausdorff (BCH) formula, we can expand the operator evolution as:
\begin{align}\label{eq:operator_evolution}
    O(t) &= O + (it)[H, O] + \frac{(it)^2}{2!}[H,[H, O]] + \frac{(it)^3}{3!}[H,[H,[H, O]]] + \dots \,,\nonumber \\
    &= \sum_{n=0}^{\infty} \frac{(it)^n}{n!} \mathcal{L}^n O \nonumber \\
    &= e^{i \mathcal{L} t} O
\end{align}
where $\mathcal{L}^n$ denotes nested applications of the Liouvillian superoperator, defined as the commutator $[H, \cdot]$. The Krylov basis is defined through the nested application of $\mathcal{L}$ to the operator $O$. The Krylov subspace, within which the operator evolves, is spanned by the set of nested commutators:
\begin{equation}\label{basis}
    \mathrm{span}\{O, \mathcal{L}O, \mathcal{L}^2 O, \mathcal{L}^3 O, \dots, \mathcal{L}^n O\}.
\end{equation}

In the Krylov basis, the operator evolution is expressed as:
\begin{equation}\label{eq:krylov_basis}
    O(t) = \sum_{n=0}^{\infty} i^n \phi_n(t) O_n,
\end{equation}
where $O_n$ are the basis operators. The Krylov operator complexity is defined as \cite{Parker_2019}:
\begin{equation} \label{comnew}
    K_O(t) = \sum_{n=0}^{\infty} n |\phi_n(t)|^2,
\end{equation}
where $\phi_n(t)$ are the Krylov wavefunctions, representing the amplitude of the operator at site $n$ along the one-dimensional Krylov chain. Thus, Krylov complexity is the expectation value of the operator’s position along this chain. These wavefunctions are obtained by solving the Schrödinger-like equation  \cite{Parker_2019, dymarsky2020quantum, Dymarsky_2021}:
\begin{equation}\label{eq:schrodinger_like}
    \partial_t \phi_n(t) = b_n \phi_{n-1}(t) - b_{n+1} \phi_{n+1}(t) - a_n \phi_n(t),
\end{equation}
where $a_n$ and $b_n$ are the Lanczos coefficients. From (\ref{eq:krylov_basis}) and (\ref{eq:schrodinger_like}), one can see that the system’s dynamics are fully encoded in the Lanczos coefficients. For unitary (closed) systems, the $a_n$ coefficients vanish. The Lanczos coefficients can be computed using methods such as the Lanczos algorithm or the moments of the autocorrelation function \cite{Parker_2019, dymarsky2020quantum, Dymarsky_2021,vish123}.
\subsection*{Moments Method}
The moments method involves computing the Lanczos coefficients via the two-point correlation function, also known as the autocorrelation function or return amplitude \cite{Dymarsky_2021, Balasubramanian:2022tpr}:
\begin{equation}
    C(t) = \langle O(t) | O(0) \rangle = \phi_0(t).
\end{equation}

The moments are obtained by taking the $n$-th time derivative of the correlation function :
\begin{equation}\label{eq:moments}
    \mu_n =  \frac{d^n}{dt^n} \langle O(t) | O(0) \rangle .
\end{equation}

The Lanczos coefficients $\{b_n\}$ and $\{a_n\}$ are related to the moments $\{\mu_n\}$ via \cite{Hashimoto_2023,Parker_2019, dymarsky2020quantum, Dymarsky_2021,vish123}:
\begin{align}
    b^2_n(t) &= \frac{\tau_{n+1}(t) \tau_{n-1}(t)}{\tau_n^2(t)} \quad n \geq 0, \\
    a_n(t) &= \frac{d}{dt} \log\left( \frac{\tau_n(t)}{\tau_{n-1}(t)} \right) \quad n \geq 0, \label{new11}
\end{align}
where $\tau_n(t)$ are determinants of matrices constructed from the moments:
\begin{equation}
    \tau_n(t) = \begin{vmatrix}
        \mu_0 & \mu_1 & \mu_2 & \cdots & \mu_n \\
        \mu_1 & \mu_2 & \mu_3 & \cdots & \mu_{n+1} \\
        \mu_2 & \mu_3 & \mu_4 & \cdots & \mu_{n+2} \\
        \vdots & \vdots & \vdots & \ddots & \vdots \\
        \mu_n & \mu_{n+1} & \mu_{n+2} & \cdots & \mu_{2n}
    \end{vmatrix},
\end{equation}
with $\tau_{-1}(t) = 0$. Typically, the coefficients are evaluated at a cutoff time $t_c = 0$, except in certain cases, such as conformal field theories (CFTs). The evolution of the operator, driven by the Liouvillian as shown in (\ref{eq:operator_evolution}), is central to studying Krylov complexity, which quantifies operator growth. This can be visualized as follows:

\begin{figure}[H]
    \centering
    \begin{tikzpicture}[scale=1.7]
    \definecolor{myred}{RGB}{255, 0, 0}
    \draw[->] (-2, 0) -- (2, 0) node[below] {Operator Space};
    \draw[->] (-1, 0) -- (-1, 3) node[anchor=south] {$t$};
    \path[fill=blue!40, opacity=0.6] (0,0) -- (-1.5,2.5) -- (1.5,2.5) -- cycle;
    \filldraw[black] (0,0) circle (2pt);
    \node at (0.5,2) {$\mathcal{O}(t)$};
    \node at (0,-0.4) {$\mathcal{O}(t=0)$};
    \end{tikzpicture}
    \caption{Schematic diagram of the spreading of the operator, as it evolves, in operator space.}
    \label{fig:OperatorGrowth}
\end{figure}

\subsection*{Krylov Complexity for Open Quantum Systems}\label{K-complexityForOpenSytems}

 \textcolor{black}{For open quantum systems governed by a Lindbladian \(\mathcal{L}\) (a generalized Liouvillian), which describes non-unitary evolution due to environmental coupling, the standard Lanczos algorithm for Hermitian operators is insufficient. Instead, there are a variety of ways in which one can treat such quantum systems using Krylov methods. Among them are the {\it Arnoldi algorithm, Closed Krylov basis} and the {\it bi-Lanczos algorithm}. {\it The closed Krylov basis} \cite{Liu_2023} is an identical treatment to the generalization of the closed system dynamics for a non-Hermitian initial operator, as was discussed in the previous sub-section. Both the Arnoldi and bi-Lanczos algorithms attempt to generalize the Lanczos algorithm for open systems. {\it The Arnoldi algorithm} does this by constructing an orthonormal basis; however, it turns out that in this basis, the Lindbladian is not tri-diagonal, instead it is in an upper Hessenberg form. On the other hand, {\it the bi-Lanczos algorithm } aims to preserve the tri-diagonal structure obtained in the Lanczos algorithm. This is done by evolving two initial vectors $|p_0 \rangle $ and $|q_0 \rangle$ by the adjoint Lindbladian $\mathcal{L^\dag}$ and the Lindbladian $\mathcal{L}\,,$ respectively. As such, it creates a bi-orthonormal basis instead of an orthonormal one,}

\begin{align}\label{BiLanczos}
    \mathrm{Kry}^j (\mathcal{L^\dag}, |p_0 \rangle) &= \{|p_0 \rangle ,\mathcal{L^\dag} |p_0 \rangle, (\mathcal{L^\dag})^2 |p_0 \rangle, \hdots \}, \\
    \mathrm{Kry}^j (\mathcal{L}, |q_0 \rangle ) &= \{|q_0 \rangle,\mathcal{L} |q_0 \rangle, (\mathcal{L})^2 |q_0 \rangle, \hdots \},
\end{align}

\textcolor{black}{with the bi-orthonormality condition, $\langle q_m | p_n \rangle = \delta_{mn}$ \cite{Bhattacharya_2023,Chakrabarti:2025hsb}. Unlike the Arnoldi algorithm, which leaves the Linbdladian in an upper Hessenberg form, the bi-Lanczos algorithm, as mentioned earlier, preserves the tri-diagonal structure with complex diagonal elements \(a_n\) and off-diagonal elements \(b_n\) (super-diagonal) and \(c_n\) (sub-diagonal). Generated by equivalent statements of (\ref{BiLanczos}), by the action of $L$, $L^{\dag}$ on the vectors.}

\begin{align}
    \mathcal{L^{\dag}}|p_j\rangle &= b_j |p_{j-1}\rangle +a_j |p_j\rangle + c_{j+1} |p_{j+1}\rangle\,, \\
    \mathcal{L}|q_j\rangle &= c^*_j |q_{j-1}\rangle +a^*_j |q_j\rangle + b^*_{j+1} |q_{j+1}\rangle\,.
\end{align}


\textcolor{black}{Using this representation, the matrix elements of the Lindbladian can be written as $\langle q_i|\mathcal{L^{\dag}}|p_j \rangle $, allowing it to be expressed as: }

\begin{equation}
    \mathcal{L^{\dag}} = \begin{pmatrix}
    a_0 & b_0 & 0 & \cdots \\
    c_0 & a_1 & b_1 & \ddots \\
    0 & c_1 & a_2 & \ddots \\
    \vdots & \ddots & \ddots & \ddots
    \end{pmatrix}
\end{equation}

\textcolor{black}{where \(a_n, b_n, c_n \in \mathbb{C}\). Using a similarity transformation, the off-diagonal entries can be written as a generalized Lanczos coefficient, $\tilde{b}_n = \sqrt{b_n c_n}$. Define a diagonal matrix \(D = \text{diag}(d_0, d_1, \dots)\) with \(d_0 = 1\) and ratios $d_k = \sqrt{c_k / b_k}$. The transformed Lindbladian is \(\mathcal{L}' = D^{-1} \mathcal{L} D\), which preserves eigenvalues and has the form \cite{Yamamoto1968}:}

\begin{equation}
    \mathcal{L}' = \begin{pmatrix}
    a_0 & \sqrt{b_0 c_0} & 0 & \cdots \\
    \sqrt{b_0 c_0} & a_1 & \sqrt{b_1 c_1} & \ddots \\
    0 & \sqrt{b_1 c_1} & a_2 & \ddots \\
    \vdots & \ddots & \ddots & \ddots
    \end{pmatrix}. \label{new1}
\end{equation}




\textcolor{black}{The benefit of this similarity transformation is that it allows us to generalize the moments method for closed systems to open systems. In general, the Lanczos coefficients $b_n$ are taken to be positive real numbers, with the diagonal coefficients generally being complex. For complex values of $\sqrt{b_n c_n}$ 
\cite{gaaf2016infinitebilanczosmethodnonlinear}, these terms can be written in modulus–argument form. In this representation, the equal off-diagonal elements share constant phases, which are absorbed by the transformation matrix $D_1$.}

\begin{equation}
    D_1 = \mathrm{diag}\left(1,\exp\left\{i\phi_1 \right\}, \exp\left\{i(\phi_1 + \phi_2 \right\}, \hdots, \exp\left\{i\sum_n \phi_n \right\} \right) 
\end{equation}

Such that, 

\begin{equation}
    \mathcal{L}'= D^{-1}_1 \mathcal{L} D_1\,.
\end{equation}

\textcolor{black}{This redefines the basis up to an overall phase, so that only the moduli of the coefficients need to be considered. This is possible as the diagonal coefficients $a_n$ are the ones that ensure the non-Hermitian nature of the Lindbladian \cite{Nandy_2025}. In the absence of the complex diagonal coefficients, the Lindbladian/Liouvillian should be hermitian with the off-diagonal coefficients being, in the most general case, complex conjugate pairs \cite{Caputa:2024vrn, Dymarsky_2021}. Note that, in some dissipative systems, such as the SYK model with Lindblad operators \cite{Bhattacharjee:2023uwx}, \(b_n = c_n\) may be real and positive, simplifying to \(\sqrt{b_n c_n} = b_n\). For general cases, the transformation symmetrizes the off-diagonals, allowing for a form of a generalized moments method for Krylov complexity, which accounts for dissipative effects, where the imaginary parts of \(a_n\) typically suppress complexity growth.}\\

\textcolor{black}{As mentioned above, in this transformed basis, the Lanczos coefficients $\{b_n\}$  are replaced by $\tilde{b}_n = \sqrt{b_n c_n}$ and are related to the moments $\{\mu_n\}$ as \cite{NSSrivatsa:2023pby}:}

\begin{align}
   \tilde{b}^2_n = b_n c_n &= \frac{\tau_{n+1}(t) \tau_{n-1}(t)}{\tau_n^2(t)}, \quad n \geq 0\,, \label{new111}
\end{align}
,\textcolor{black}{and the $a_n' s$ are still computed using the definition given in (\ref{new11}). In the following, we first consider a damped harmonic oscillator, followed by the well-known Caldeira–Leggett model as our representative open quantum systems. For both models, the Krylov basis is not large, and the $b_n$ Lanczos coefficients truncate fairly quickly, as is the case for the Harmonic Oscillator (shown in Appendix~(\ref{AppA})). The Hilbert space of the system is infinite-dimensional, but because of the choice of the initial operator and associated algebra, the Lanczos $b_n$ coefficients terminate rapidly, as we will show later in Fig.~(\ref {fig:LanczosCoeff}). Once we have the Lindbladian in tri-diagonal form as shown in (\ref{new1}), we can use (\ref{eq:schrodinger_like}) to compute the Krylov elements $\phi_n's$ and study the evolution of Krylov complexity (defined in (\ref{comnew})) in this primed basis. Also, we will show that $\tilde{b}_n's$ mentioned above will turn out to be positive real numbers, which will serve as a consistency check of the above discussion. }

\section{Damped Harmonic Oscillator}
\label{DampedOsc}
The first model we consider is the well-known damped harmonic oscillator. Its Hamiltonian describes a harmonic oscillator coupled to an electromagnetic field, serving as a representative example of damping:
\begin{equation}
    H = \omega_1 a^\dagger a + \sum_k \omega_k b_k^\dagger b_k + \sum_k \gamma_k (a^\dagger b_k + a b_k^\dagger).
\end{equation}
Unlike closed quantum systems, open quantum systems do not evolve unitarily. Instead, their dynamics are governed by equations of motion for the density matrix, which generalize classical Markovian processes. This formalism leads to the Lindblad master equation:
\begin{equation}\label{eq:lindblad}
    \frac{d \rho_s(t)}{dt} = -i [H, \rho_s(t)] + \sum_i \left( L_i^\dagger \rho_s(t) L_i - \frac{1}{2} L_i^\dagger L_i \rho_s(t) - \frac{1}{2} \rho_s(t) L_i^\dagger L_i \right).
\end{equation}
This equation assumes a Markovian bath, where bath correlations are proportional to delta functions in time, implying a memoryless environment where information lost to the bath cannot return to the system. This is central to dissipative 
dynamics. \textcolor{black}{We know that in both the Schrodinger and Heisenberg picture, the expectation value of some operator $O$ remains the same, i.e, \begin{align}
    \Tr{O(0) \rho_s(t)} &= \Tr{O(t) \rho_s(0)}\,, \nonumber\\
    \Tr{O(0) \frac{d}{dt} \rho_S(t)} &= \Tr{\frac{d}{dt}O(t) \rho_s(0)}\,,\nonumber\\
    \Tr{O(0) \mathcal{L} \rho_s(t)} &= \Tr{\mathcal{L}^{\dag}O(t) \rho_s(0)}\,.
\end{align}
Hence, the operator evolution equation must be, 
\begin{equation}
    \frac{dO(t)}{dt} = \mathcal{L}^{\dag} O(t).
\end{equation}
Making use of the definition of $\mathcal{L}$ and the cyclic nature of the trace, one arrives at the following adjoint master equation, describing operator evolution:
\begin{equation}
    \frac{dO(t)}{dt} = i [H, O(t)] + \sum_i \left( L_i^\dagger O(t) L_i - \frac{1}{2} L_i^\dagger L_i O(t) - \frac{1}{2} O(t) L_i^\dagger L_i \right)\,.
\end{equation}
The first term generates unitary evolution, while the remaining terms account for transitions due to interactions with the bath. The operators $L_i$ are jump operators, with $L_i^\dagger A L_i$ driving quantum transitions and the terms $-\frac{1}{2} L_i^\dagger L_i O(t)$ and $-\frac{1}{2} O(t) L_i^\dagger L_i$ correcting for cases where no jump occurs \cite{Preskill1998}.}\\

\begin{figure}[htb!]
\centering
\includegraphics[width=.65\textwidth]{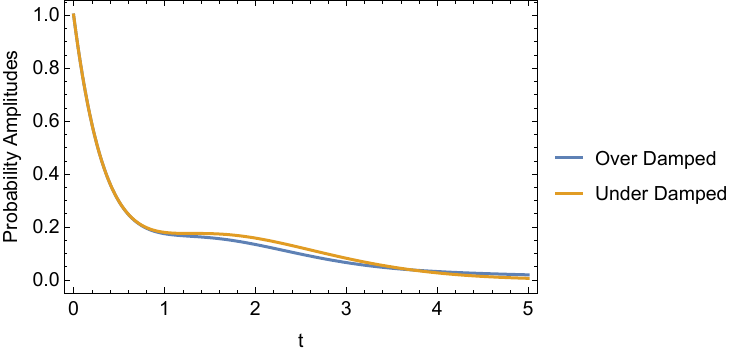}
\qquad
\caption{Time evolution of probability amplitudes for under-damped ($\omega = 1\,,\gamma = 0.1$) and over-damped ($\omega = 1\,,\gamma = 0.9$) systems.\label{fig:ii}}
\end{figure}

Using the framework mentioned in Sec.~(\ref{KryComp}), we study the Krylov complexity in this dissipative system. The jump operators are chosen to maintain the Markovian approximation. In open systems, information and energy flow bidirectionally between the system and bath, leading to non-Markovian fluctuations. A good Markovian approximation requires sufficient separation between the correlation time of fluctuations and the system’s evolution timescale. We model the system-bath interaction as:
\begin{equation}
    H_I = \sum_k \gamma_k (a b_k^\dagger + a^\dagger b_k)\,.
\end{equation}
Assuming the environment is initially at zero temperature, the system can emit energy without absorbing any from the bath. We define a single Lindblad operator for photon emission, proportional to the annihilation operator $a$:
\begin{equation}
    L = \sqrt{\gamma} a\,,
\end{equation}
where $\gamma$ is the decay rate from the first excited state to the ground state. The master equation for a system operator $O(t)$ becomes:
\begin{equation}
    \dot{O}(t) = i [H_0, O(t)] + \gamma \left( a^\dagger O(t) a - \frac{1}{2} a^\dagger a O(t) - \frac{1}{2} O(t) a^\dagger a \right)
\end{equation}
where $H_0 = \omega_1 a^\dagger a$ is the standard harmonic oscillator Hamiltonian. For the Caldeira-Leggett model, we will consider a more generic quadratic form for the operator $O = \frac{1}{2\omega} (a^\dagger a^\dagger + a a + a^\dagger a + a a^\dagger)$, which then evolves as:
\begin{equation}
    O(t) = \frac{1}{2\omega} \left( f_1(t) a^\dagger a^\dagger + f_2(t) a a + f_3(t) a^\dagger a + f_4(t) a a^\dagger \right),
\end{equation}
where $\{f_1, f_2, f_3, f_4\} \in \mathbb{C}$. Solving the differential equation for the individual annihilation and creation operators $a(t)$ and $a^{\dag}(t)$, we obtain:
\begin{align}\label{CreAnn}
a^{\dag}(t) &= e^{i \omega t - \frac{\gamma }{2}t } a^{\dag}\,,\\
    a(t) &= e^{-i \omega t - \frac{\gamma }{2}t } a \,.
\end{align}
Combining these yields the following:
\begin{equation}
    O(t) = \frac{1}{2 \omega}(e^{-i 2\omega t - \gamma t} a a + e^{i 2\omega t - \gamma t} a^\dagger a^\dagger + e^{- \gamma t} a^\dagger a +  e^{-\gamma t} a a^\dagger)\,.
\end{equation}
This result is consistent with the harmonic oscillator, with additional exponential decay terms reflecting information loss to the environment. The thermal two-point correlation function is:
\begin{equation}
    \langle Q(t) Q(0) \rangle_{\beta \rightarrow \infty} = 2 e^{-i 2\omega t - \gamma t} - e^{- \gamma t}\,.
\end{equation}
Using the moments method from Sec.~(\ref{KryComp}) \textcolor{black}{(using equations (\ref{new11}) and (\ref{new111}))}, we compute the probability amplitudes $(|\phi_n|^2)$ and the corresponding Krylov complexity for the under and overdamped cases, which are displayed in Fig.~(\ref{fig:i}) and (\ref{fig:ii}) respectively for different values of $\omega$: 

\begin{figure}[ht!]
\centering
\includegraphics[width=.65\textwidth]{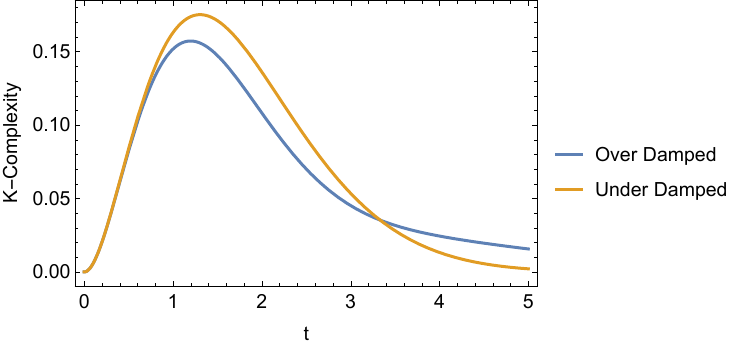}
\qquad
\caption{Time evolution of Krylov complexity for under-damped ($\omega = 1\,, \gamma = 0.1$) and over-damped systems ($\omega = 1\,,\gamma = 0.9$).\label{fig:i}}
\end{figure}

From the Fig.~(\ref{fig:i}), we observe that at early times Krylov complexity exhibits rapid growth, followed by a decay toward a low saturation value. This behavior contrasts with the harmonic oscillator results presented in the Fig.~(\ref{fig:A2}) of the Appendix~(\ref{AppA}). Unlike the harmonic oscillator, which is a closed system, the damped oscillator loses information to the environment without any backflow. This is captured in the presence of the complex $a_n$ Lanczos coefficients, as for the harmonic oscillator and the damped oscillator, there is only one non-zero $b_n$ coefficient. For the harmonic oscillator, the complexity displays periodic oscillations between zero and one, reflecting the evolution of the position operator into the momentum operator and vice versa, as dictated by the canonical commutation relations. These periodic oscillations are absent in the damped case.

In line with the conjecture made in \cite{Liu_2023}, dissipation leads to a lower overall complexity and a reduced saturation value. We therefore identify the oscillatory decay of Krylov complexity to a low saturation value as a hallmark of dissipation. Furthermore, the maximum complexity reached is significantly smaller than in the harmonic oscillator, as can be seen from the Fig.~(\ref{fig:i}) and (\ref{fig:A2}) of the Appendix~(\ref{AppA}). This suggests that, in the overdamped regime, the operator explores only a limited portion of the system space before dissipating into the bath, while in the underdamped regime, it explores more of the system space (compared to the overdamped case, as evidenced by its higher saturation value) prior to this loss. Nevertheless, in both cases the spread into the bath occurs on the same timescale, consistent with the observed saturation. This interpretation is further supported by the probability amplitude plots as shown in Fig.~(\ref{fig:ii}). For the underdamped case, the slower decay indicates a less rapid loss of information to the bath.

\section{Caldeira-Leggett master equation}
\label{CLmod}
 The Caldeira-Leggett model is a system-bath model designed to study quantum Brownian motion \cite{CALDEIRA1983374, Caldeira:1982iu, PhysRevLett.46.211}. It describes a Brownian particle of mass $m$ coupled via coordinate-coordinate coupling to a bath of non-interacting harmonic oscillators:

\begin{equation}
    H = \underbrace{\frac{p^2}{2m} + \frac{m \omega^2 x^2}{2}}_{H_s} + \sum_n \left( \frac{p_n^2}{2m_n} + \frac{m_n \omega_n^2 x_n^2}{2} \right) + x \sum_n \kappa_n x_n\,.
\end{equation}
In the Schrödinger picture, the dynamics of the reduced system (i.e for the particle of mass $m$) are described by the Caldeira-Leggett master equation:

\begin{equation}\label{eq:cl_master}
    \frac{d}{dt} \rho_s(t) = -\frac{i}{\hbar} [H_s, \rho_s(t)] - \frac{i \gamma}{\hbar} [x, \{p, \rho_s(t)\}] - \frac{2 m \gamma k_B T}{\hbar^2} [x, [x, \rho_s(t)]]\,.
\end{equation}

Here the subscript $s$ denote the reduced system obtained by tracing out the bath. As seen from (\ref{eq:cl_master}), the terms in the master equation come directly from the dissipative and noise kernels respectively, as such the first term describes \textit{coherent dynamics}, the second term, proportional to the \textit{relaxation rate} $\gamma$, drives \textit{dissipation} by damping momentum expectation values, and the third term, proportional to temperature $T$, induces \textit{decoherence} through thermal fluctuations \cite{Schlosshauer_2019, BRE02}. The Caldeira-Leggett (C-L) master equation is not directly in Lindblad form, as its dissipator is not positive definite. In the high-temperature limit, this can be remedied by adding a minimally invasive term, as was seen in the discussion in the derivation of the C-L master equation
\cite{Ramazanoglu_2009,BRE02,Schlosshauer_2019}.
The Lindbladian form facilitates analysis of operator dynamics in the adjoint form:

\begin{equation}\label{LindbladCL}
    \frac{d}{dt} O(t) = i \left [ H_s + \frac{\gamma}{2} (x p + p x), O(t) \right] + \gamma \left( L^\dagger O(t) L - \frac{1}{2} L^\dagger L O(t) - \frac{1}{2} O(t) L^\dagger L \right),
\end{equation}

with the Lindblad jump operator defined as

\begin{equation} \label{lind}
    L = A x + i B p
\end{equation}

where the coefficients $A$ and $B$ are the following: 

\begin{align}
    A = \sqrt{\frac{4 m k_B T}{\hbar^2}}\,,\quad
    B = \sqrt{\frac{1}{4 m k_B T}}\,.
\end{align}

In a similar manner to how we computed the operator evolution in the case of the damped oscillator. We begin with an {\bf initial operator} $O(0) = p^2$, then the commutator structure of the RHS of the master equation (\ref{LindbladCL}) yields the following operator expression \footnote{The proportionality sign is there to indicate that the RHS of this expression does not have any time-dependent contribution.}:

\begin{align}
    \frac{d}{dt} O (t) &\propto -\omega^2 (x p + p x) - 2 \gamma p^2 + \gamma (A^2 - 2 A B p^2)\,,
\end{align}

Gathering terms and writing the identity term in terms of creation and annihilation operators, the expression can be written as

\begin{align}
    \frac{d}{dt} O(t)
    &\propto -\omega^2 (x p + p x) + (-2 \gamma - 2 \gamma A B) p^2 + \gamma A^2 (a a^\dagger - a^\dagger a)\,. \label{maineq}
\end{align}
 
Details of the derivation are given in Appendix~(\ref{AppB}).  Solving the differential equation and gathering the terms together, we obtain the following time-evolved operator

\begin{align}
    O (t) &= -\frac{\omega}{2} \Big( e^{(2 i \omega - 2 \gamma - 2 \gamma A B) t} a a - e^{(\gamma A^2 - 2 \gamma - 2 \gamma A B) t} a a^\dagger \notag\\ 
    &- e^{(-\gamma A^2 - 2 \gamma - 2 \gamma A B) t} a^\dagger a
    + e^{(-2 i \omega - 2 \gamma - 2 \gamma A B) t} a^\dagger a^\dagger \Big)\,.
\end{align}

Assuming the system is in thermal equilibrium, the density matrix $\rho_{\beta}$ is taken to be a thermal Gibbs state of the form $\rho_\beta = \frac{e^{-\beta H}}{Z}$. The auto-correlation function with respect to the thermal average is then given by:

\begin{align}
    \langle O(t) O \rangle_{\beta} &= \frac{\mathrm{Tr} \{ \rho_\beta O(t) O \}}{\mathrm{Tr} \{ \rho_\beta O^2 \}}\,, \\
    &= e^{(2 i \omega - 2 \gamma - 2 \gamma A B) t} ( C_3 + 3  C_2 + 2  C_1) + e^{(\gamma A^2 - 2 \gamma - 2 \gamma A B) t} (2  C_3 + 3  C_2 +  C_1) \notag \\
    & \quad + e^{(-\gamma A^2 - 2 \gamma - 2 \gamma A B) t} (2  C_3 +  C_2) + e^{(-2 i \omega - 2 \gamma - 2 \gamma A B) t} ( C_3 -  C_2)
\end{align}

where the coefficients $C_1, C_2$ and $C_3$ are given as,

\begin{align} 
    C_1 &= \frac{1}{\mathrm{Tr}\{\rho_{\beta} O^2\}}\sum_n e^{-\beta(n+1/2)\omega} = \frac{(-1+e^{\beta \omega})^2}{3(e^{\beta \omega}+1)^2}\,,\nonumber \\
    C_2 &= \frac{1}{\mathrm{Tr}\{\rho_{\beta} O^2\}}\sum_n n e^{-\beta(n+1/2)\omega} = \frac{-1+e^{\beta \omega}}{3(e^{\beta \omega}+1)^2}\,, \label{eq2}\\
    C_3 &= \frac{1}{\mathrm{Tr}\{\rho_{\beta} O^2\}}\sum_n n^2 e^{-\beta(n+1/2)\omega} = \frac{1}{3+3e^{\beta \omega }}\,.\nonumber
\end{align}

These coefficients and the two-point function are evaluated at various temperatures, consistent with the high-temperature limit of the Caldeira–Leggett master equation.\\

To investigate the role of dissipation and correlations, we begin in the following subsections with an analysis of the decoherence function, followed by a study of Krylov complexity. Our aim is to determine whether Krylov complexity is sensitive to decoherence in a manner similar to circuit complexity \cite{Bhattacharyya_2024,Bhattacharyya_2023}.

\subsection{Decoherence Function}  \label{newsec22}
Interactions between an open quantum system and its environment create correlations between states of the system and the environment. In a specific basis these correlations/coherence in the density matrix are strongly decaying, a process tracked by the decoherence function, which describes the behavior of the off-diagonal elements of the reduced density matrix. Consider, as the starting point of our discussion, the Hamiltonian:
\begin{equation}
    H = H_S + H_B + H_I = H_0 + H_I,
\end{equation}
The interaction Hamiltonian takes the following form,
\begin{equation}
    H_I = \sum_n |n \rangle \langle n | \otimes B_n \equiv \sum_n A_n \otimes B_n.
\end{equation}
This interaction Hamiltonian singles out a specific set of basis vectors of the system, and $B_n$ denotes reservoir operators. The choice of $A_n = |n \rangle \langle n|$ is such that $A_n$ are conserved quantities, i.e., that 
\begin{equation}\label{BasisCondition}
    [H_0 + H_I, A_n]= [H_0, A_n] = 0\,.
\end{equation}
By this condition, it is clear that the Krylov basis is not a suitable basis for deriving the decoherence function. As a requirement of the Lanczos algorithm,  the commutation relations between the Hamiltonian and the operator are non-trivial; otherwise, (\ref{basis}) would terminate instantly. Then, from (\ref{BasisCondition}), we can see that in the interaction picture, the time evolution of the interaction Hamiltonian passes entirely onto the bath operators. 
\begin{equation}
    H_I(t) = e^{i H_0 t} H_I e^{-i H_0 t} = \sum_n |n \rangle \langle n | \otimes B_n(t) \,.
\end{equation}
As a result of the commutation relation, the basis states $|n\rangle$ are not affected by the coupled dynamics. This gives us the following system bath state:
\begin{equation}
    |\Psi(t)\rangle = \sum_n c_n |n\rangle \otimes |\phi_n(t)\rangle\,.
\end{equation}
It represents correlations between states describing the system and the corresponding states of the reservoir. Due to the interaction, the reservoir carries information on the system state. These coherence/correlations are still present in the system's reduced density matrix:
\begin{equation}\label{reducedEq}
    \rho_s(t) = \mathrm{Tr}_B{|\Psi(t)\rangle \langle\Psi(t)|} = \sum_n \sum_m c_n c^{*}_m |n\rangle \langle m| \langle \phi_m(t)|\phi_n(t)\rangle\,.
\end{equation}
It is clear that the diagonal elements of $\rho_s(t)$ are constant in time while the off-diagonals vary in time. The time dependence of $\rho_s(t)$ is given by the overlap between the corresponding reservoir states: 
\begin{equation}
    \langle \phi_m(t)|\phi_n(t)\rangle = \exp{[\Gamma_{n m}(t)]}, \hspace{0.5cm} \Gamma_{n m}(t) \leq 0\,.
\end{equation}
The quantity $\Gamma_{n m}(t)$ characterizes the behaviour of the off-diagonal terms in the reduced density matrix and is known as the decoherence function. For the Caldeira-Leggett model, the decoherence function is \cite{BRE02}:
\begin{align}
    \Gamma(t) &= -\frac{\Delta x^2 + 4 \sigma_0^4 \Delta p^2}{8 \sigma_0^2} \frac{\langle I^2(t) \rangle}{\langle I^2(t) \rangle + \sigma_0^2 G_1^2(t) + \frac{G_2^2(t)}{4 m^2 \sigma_0^2}}\,,\nonumber \\
    \Tilde{\Gamma}(t) &= \frac{8 \sigma_0^2 \Gamma(t)}{\Delta x^2 + 4 \sigma_0^4 \Delta p^2} = \frac{\langle I^2(t) \rangle}{\langle I^2(t) \rangle + \sigma_0^2 G_1^2(t) + \frac{G_2^2(t)}{4 m^2 \sigma_0^2}}
\end{align}
where $I$, $G_1$, and $G_2$ are the inhomogeneous and homogeneous solutions of the operator $x(t)$ evolution, and $\Delta x$ and $\Delta p$ are the initial separations of superposed Gaussian wave packets in position and momentum space, respectively.
\begin{figure}[H]
\centering
\includegraphics[width=.55\textwidth]{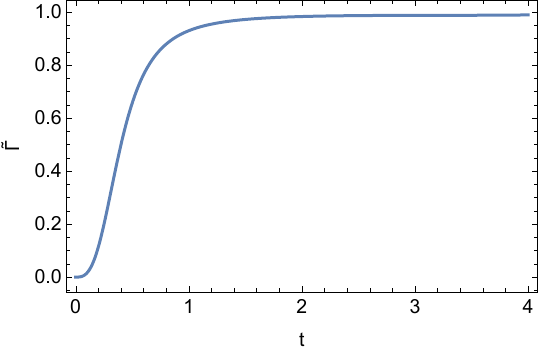}
\caption{Time evolution of decoherence function $\Tilde{\Gamma}$ for $\omega = 1$ and $\gamma = 0.1.$ \textcolor{red}. \label{fig:3}}
\end{figure}
From Fig.~(\ref{fig:3}), it is clear that the decoherence rate saturates, as described in \cite{Schlosshauer_2019}. This occurs when the separation between two system positions $x$ and $x'$ reaches the maximum coherence length of the environment\footnote{ When the separation between $x$ and $x'$ exceeds the coherence length, as the environment is performing a type of ``measurement" on the system, it cannot distinguish any further between the points, causing decoherence to saturate.}  The saturation of the decoherence function implies that the off-diagonal elements of the density matrix vanish. However, this feature is basis-dependent, and as noted earlier in the section, the Krylov basis is not the appropriate one for capturing it. We therefore suspect that Krylov complexity may be insensitive to decoherence, a possibility we will explore in the next subsection.

\subsection{Lanczos Coefficients and Krylov Complexity} \label{secnew1}

\begin{figure}[H]
\centering
\includegraphics[width=.55\textwidth]{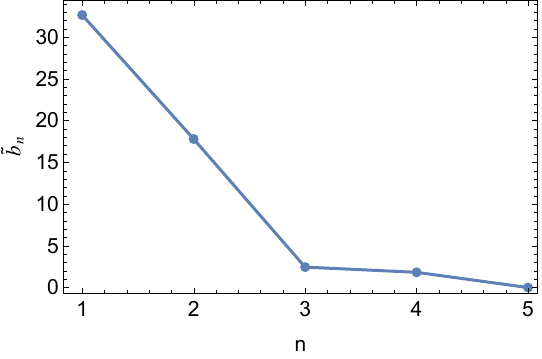}
\caption{$\tilde{b}_n$ Lanczos coefficients for the C-L model, with $\omega = 1$ and $\gamma = 1/10\,.$ \label{fig:LanczosCoeff}}
\end{figure}

Now we will turn our attention to the Krylov complexity. Before presenting the results for it, we will first examine the nature of the Lanczos coefficients, which subsequently enter into the computation of the complexity. We have used the moment methods outlined in Sec~(\ref{KryComp}) to compute them using (\ref{new11}) and (\ref{new111}). Note that the dimension of the Krylov space largely depends on the underlying algebra and the choice of the initial operator, but \textcolor{black}{does not depend on the choice of parameter in the master equation. The Krylov dimension when we consider the full C-L master equation is $\dim[\mathrm{Krylov}]=5\,.$} For the cases considered in this paper, we start with a simple operator to probe the effects of dissipation and decoherence in these models. Fig.~(\ref{fig:LanczosCoeff}) is a plot of the Lanczos coefficients $\tilde{b}_n$ vs $n\,.$ \textcolor{black}{Note that as discussed in Sec.~(\ref{KryComp}), $\tilde{b}_n's$ are real, showing the self-consistency of our method}. 

\begin{figure}[H]
\centering
\includegraphics[width=.55\textwidth]{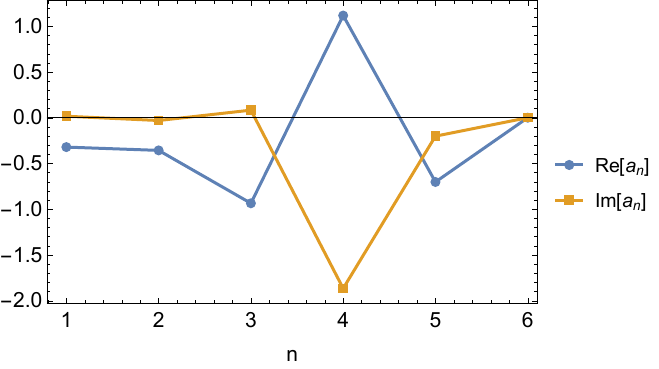}
\caption{$a_n$ Lanczos coefficients for the C-L model, with $\omega = 1$ and $\gamma = 1/10$. \label{fig:LanczosCoeff1}}
\end{figure}

Similar to the results for the closed system (the quantum harmonic oscillator), and as expected, the Lanczos coefficients terminate quickly as a result of the algebra and initial operator choice. \textcolor{black}{The number of Lanczos coefficients and thus the Krylov dimension does not depend on the parameters $(T, \omega, \gamma)$, adjusting the parameters merely scales the coefficients as demonstrated in Appendix~(\ref{appc}).} While the $a_n$ coefficients are complex (as shown in Fig.~(\ref{fig:LanczosCoeff1})) and account for the dissipative features in the Krylov complexity, the ${\tilde{b}_{n}}'s$ remain real. 

\begin{figure}[H]
\centering
\includegraphics[width=.45\textwidth]{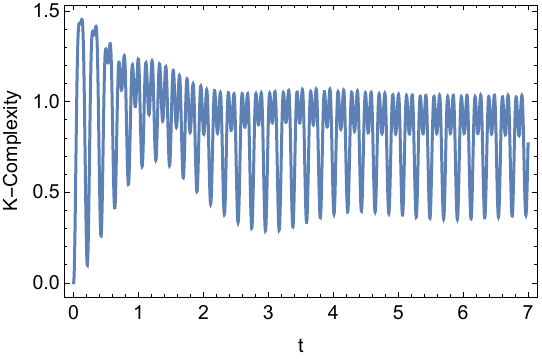}
\qquad
\includegraphics[width=.45\textwidth]{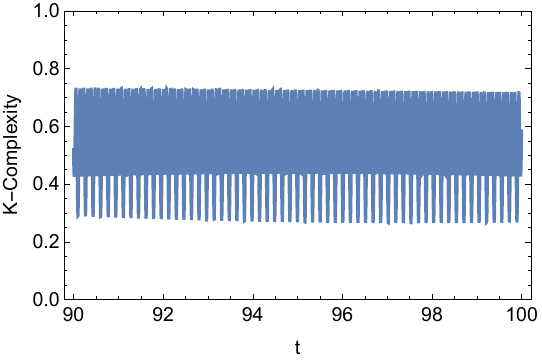}
\caption{Time evolution of Krylov complexity 
for the Caldeira-Leggett master equation with $\omega = 1$ and $\gamma = 1/10\,.$ On the right, the late time complexity for the same values of $\gamma$ and $\omega$ is displayed. }\label{fig:iii}
\end{figure}

It is clear that, simply by examining the behavior of the Lanczos coefficients, we cannot obtain any significant information about the dynamics, especially regarding the role of dissipation. Hence, we turn our attention to the Krylov complexity. As outlined in Sec.~(\ref{KryComp}), we use the moments method to compute the Krylov complexity for the Caldeira–Leggett model. To investigate the roles of dissipation and correlation, we examine the time evolution of the Krylov complexity, shown in Fig.~(\ref{fig:iii}). \textcolor{black}{The Krylov elements $\phi_n's$ needed for computing the complexity using (\ref{comnew}) are shown in Appendix~(\ref{appD})}.

\begin{figure}[H]
\centering
\includegraphics[width=.45\textwidth]{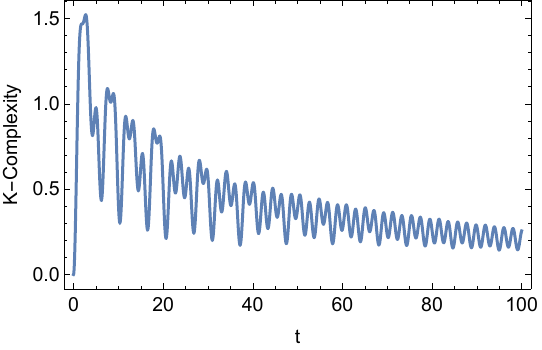}
\qquad
\includegraphics[width=.45\textwidth]{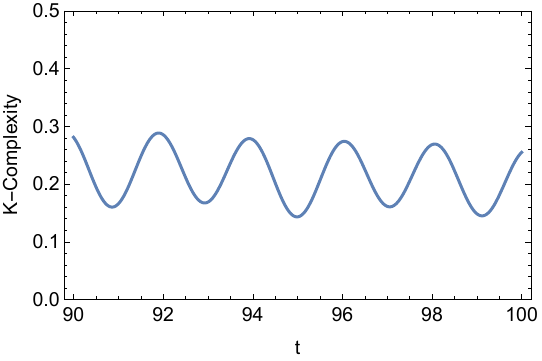}
\caption{Time evolution of Krylov complexity for the Caldeira-Leggett master equation with $\omega = 1/10$ and $\gamma = 1/20$. On the right, the saturation of the complexity for the same values of $\gamma$ and $\omega$ is displayed.\label{fig:iv}}
\end{figure}

From Fig.~(\ref{fig:iii}), we observe two key features of the Krylov complexity in the Caldeira-Leggett model:
\begin{itemize}
    \item The initial oscillatory behavior is accompanied by a secondary frequency mode—‘anti-oscillations,’ which appear at the peaks of each oscillation, as shown in Fig.~(\ref{fig:iii}) and Fig.~(\ref{fig:iv}). This feature causes the Krylov complexity to behave differently from the results in Sec.~(\ref{DampedOsc}) with dissipation. In the damped oscillator case studied in Sec.~(\ref{DampedOsc}), the contribution from decoherence was suppressed, unlike in the Caldeira–Leggett model. As a result, the Krylov complexity here exhibits a much slower decay compared to Sec.~(\ref{DampedOsc}).
    \item The complexity saturates, a well-known result for open quantum systems \cite{Bhattacharya_2023, Liu_2023}. This saturation reflects the interplay between the decoherence and dissipative terms in the master equation. 
\end{itemize}
For a given set of parameters, we find that the complexity saturates at a value close to $0.2$, within an interval of $0.1$, as illustrated in Fig.~(\ref{fig:iv}). This saturation occurs at approximately $t = 93$. The saturation is driven by the interplay of dissipation and decoherence. To confirm this, we further investigate the model in the following analysis.

\subsection{Decomposing the Master Equation into Individual Terms} \label{secnew}

Analyzing the master equation, it is clear that under particular operator choices and assumptions, we can isolate the contributions from individual terms in the master equation and study how these terms contribute to Krylov complexity. Therefore, in the following sections, we will consider particular choices that allow us to investigate the individual contributions of decoherence and dissipation. 
\subsubsection*{Decoupling Dissipation from the Master Equation}
To separate the effects of dissipation on Krylov complexity from the master equation, we turn off the term proportional to the relaxation rate $\gamma$, thereby considering the Lindblad double-commutator form of the master equation \cite {Schlosshauer_2019,BRE02}\,.
\begin{equation}\label{doublecomm}
    \frac{d}{dt} \rho_s(t) = -\frac{i}{\hbar} [H_s, \rho_s(t)]  - \frac{2 m \gamma k_B T}{\hbar^2} [x, [x, \rho_s(t)]]. 
\end{equation}
This yields a master equation that accounts only for the coherent system dynamics and the environment-induced decoherence, as discussed in \cite{Ferialdi_2017, BRE02, Schlosshauer_2019}. Using the same operator choice (i.e $O(0) = p^2$) as in Sec. (\ref{CLmod}), we compute the two-point correlation function:
\begin{align}
    \langle O(t) O(t) \rangle_{\beta}
    &= e^{-2 i \omega t} ( C_3 + 3  C_2 + 2  C_1) + e^{(-8m \gamma k_B T/\hbar^2) t} (2  C_3 + 3  C_2 +  C_1) \notag \\
    & \quad + e^{(8m \gamma k_B T/\hbar^2) t} (2  C_3 +  C_2) +  e^{2 i \omega t} ( C_3 -  C_2)
\end{align}
Using the correlation function, we compute the Krylov complexity, which only consists of the coherent system dynamics and thermal decoherence term, as is displayed in Fig.~(\ref{fig:v}) below:

\begin{figure}[H]
\centering
\includegraphics[width=.47\textwidth]{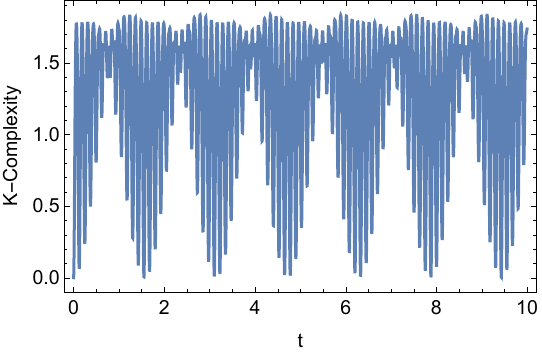}
\qquad
\includegraphics[width=.47\textwidth]{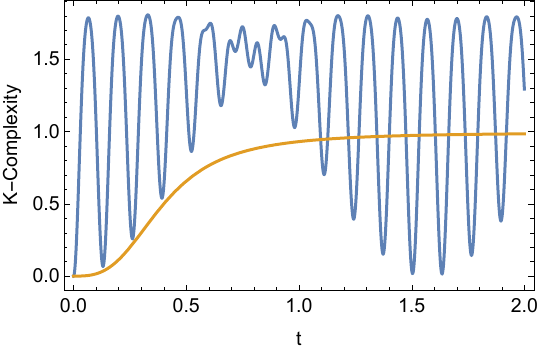}
\caption{Left: Time evolution of Krylov complexity  for $\omega = 1$ and $\gamma = 1/10$. Right: Time evolution of Krylov complexity for the same parameters, with the decoherence function $\tilde{\Gamma}(t)$ overlaid in orange.}\label{fig:v}
\end{figure}
One significant difference we immediately notice from Fig.~(\ref{fig:v}), is the absence of saturation. Note that we read off the saturation timescale from the decoherence function, and at the same timescale, the Krylov complexity, computed for identical parameters, shows diminishing oscillations. However, since the pattern remains periodic, it cannot be taken as a definitive signature of the onset of decoherence.
The varying saturation timescales in Fig.~(\ref{fig:iii}) and Fig.~(\ref{fig:iv}) arise from the competition between the dissipative and decoherence terms. In the case where the damping coefficient $\gamma$ is large compared to $\omega$ (Fig.~(\ref{fig:iv})), the saturation timescale is significantly shorter. The decoherence term generates the secondary oscillation (the ``anti-oscillations'' near each of the peaks) as observed in Fig.(\ref{fig:iv}).  We further checked that as we increased the temperature, the period of these oscillations in the Krylov complexity became shorter, appearing as ``noise’’. \\\\
Our analysis of complexity shows no clear evidence for the decay of system–bath correlations. Note that in the Krylov basis, the Lindbladian takes a tridiagonal structure characterized by the Lanczos coefficients.

\begin{equation}
    \frac{d}{dt}\rho_s(t) = \underbrace{\begin{pmatrix}
    a_0 & \sqrt{b_0 c_0} & 0 & \cdots \\
    \sqrt{b_0 c_0} & a_1 & \sqrt{b_1 c_1} & \ddots \\
    0 & \sqrt{b_1 c_1} & a_2 & \ddots \\
    \vdots & \ddots & \ddots & \ddots
    \end{pmatrix}.}_{\mathcal{L}} \rho_s(t),
\end{equation}

For $\rho_s(t)$ in the form given by (\ref{reducedEq}), it is evident that at least one of the diagonal elements of the reduced density matrix is time dependent, while at least one of the off-diagonal elements contains a term without time dependence. Consequently, there is no clear separation of the correlations carried by the off-diagonal terms, and the Krylov complexity cannot be reliably used to indicate the decay of coherence.\\

Finally, the inclusion of the dissipative term determines how rapidly the complexity decays or grows. As shown in Fig.~(\ref{fig:v}), for the same set of parameters, the complexity exhibits significantly different global features compared to Fig.~(\ref{fig:iv}).

\subsubsection*{Decoupling Decoherence}
To eliminate the effects of decoherence, we select a suitable operator that removes the temperature-dependent contribution. In particular, by choosing $O = x^2$, as in Sec.~(\ref{DampedOsc}), the decoherence term in (\ref{eq:cl_master}) is suppressed. The two-point function then becomes:
\begin{align}
    \langle O(t) O \rangle_{\beta} &= \left( e^{(4 i \omega - 2 \gamma - 2 \gamma A B) t} (C_3 + 3 C_2 + 2 C_1) + e^{(2 \omega \gamma B^2 - 2 \gamma - 2 \gamma A B) t} (2 C_3 + 3 C_2 + C_1) \right. \notag \\
    &\quad \left. + e^{(-2 \omega \gamma B^2 - 2 \gamma - 2 \gamma A B) t} (2 C_3 + C_2) + e^{(-4 i \omega - 2 \gamma - 2 \gamma A B) t} (C_3 - C_2) \right).
\end{align}
This operator choice underscores the role of the Lindblad jump operator coefficient $B$, which is proportional to the momentum, in governing the system's dissipative dynamics. In the high-temperature limit, the $B^2$ terms are suppressed, recovering the dissipative behavior discussed in Sec.~(\ref{DampedOsc}), characterized by sharp complexity growth followed by decaying oscillations toward a small saturation value. The decoherence term, however, introduces noise that manifests as oscillations around the saturation point.

\begin{figure}[htb!]
\centering
\includegraphics[width=.46\textwidth]{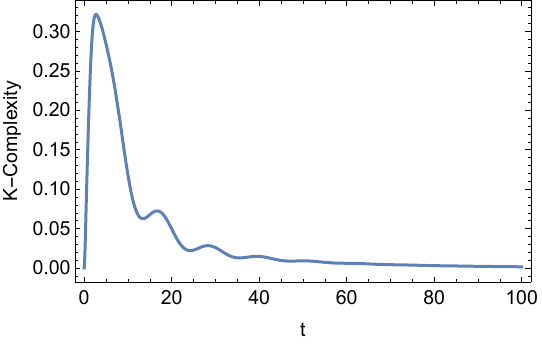}
\qquad
\includegraphics[width=.46\textwidth]{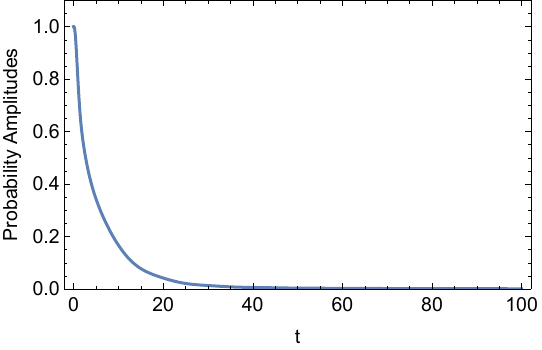}
\caption{Time evolution of Krylov complexity and probability amplitudes for the Caldeira-Leggett master equation with operator $O = x^2$, $\omega = 1$, and $\gamma = 1/10$. \label{fig:vi}}
\end{figure}
\textcolor{black}{\subsection{Krylov Complexity of density matrices in the C-L model}\label{4.4.4}
We observed in Sec.~(\ref{secnew}), Krylov complexity is unable to detect the onset of the decoherence. We now turn our attention to the Krylov complexity for the \textbf{density operator} evolved by the C-L master equation. As we are mainly interested in investigating the effect of the decoherence, we will present the study here by keeping only double commutator term in the master equation (\ref{doublecomm}) working in the regime where the decoherence dominates (readers are referred to Appendix~(\ref{AppCnew}) for a general analysis where we have kept and all the terms in the master equation and sketched the computation of the moments for the density-density correlator). In the position representation, the master equation becomes (following from (\ref{eq:cl_master})),
\begin{align}\label{positionbasis}
    \frac{\partial \rho_s(x, x', t)}{\partial t} = \left(\frac{i}{2}\left( \frac{\partial^2}{\partial x^2} - \frac{\partial^2}{\partial x'^2}\right) -  \frac{i \omega^2}{2} (x^2 - x'^2) - 2 \gamma T(x - x')^2 \right) \rho_s(x, x', t)\,.
\end{align}
\textcolor{black}{As an initial condition, we consider a Gaussian state, hence at $t=0$ the density matrix takes the following form,}
\begin{equation}
   \rho_s(x, x', 0) = \frac{1}{\sqrt{2 \pi \sigma^2}} \exp{-\frac{x^2 + x'^2}{4 \sigma^2}} \label{densitynew1}
\end{equation}
\textcolor{black}{with $\sigma = \sqrt{\hbar/2 m \omega}$. To solve this differential equation, we employ the Wigner function formalism. The density matrix can be written as a Wigner function through the transformation \cite{HILLERY1984121},}
\begin{align} 
    W(x, p) &= \frac{1}{2 \pi \hbar} \int^{\infty}_{-\infty} \left \langle x + \frac{1}{2}y \right| \rho_S \left | x - \frac{1}{2}y \right \rangle e^{i p y/\hbar} dy, \nonumber\\ 
    &= \frac{1}{2 \pi \hbar} \int^{\infty}_{-\infty}  e^{i p y/\hbar} \rho_s\left(x + \frac{1}{2}y, x - \frac{1}{2}y \right)   dy\,.\label{fourden}
\end{align}
\textcolor{black}{By making use of the relative ($r=x-x'$) and centre-of-mass co-ordinates ($X=\frac{x+x'}{2}$) and then using the  Wigner transformation, it is easy to see that the differential equation in  (\ref{positionbasis}) becomes \cite{Kramers:1940zz}, 
\begin{equation}\label{Wignerdiff}
    \frac{\partial W(X, p, t)}{\partial t} = - p\, \frac{\partial W(X, p, t)}{\partial X} + \omega^2\, X\, \frac{\partial W(X, p, t)}{\partial p} + 2 \gamma T \frac{\partial^2 W(X, p, t)}{\partial p^2}\,. 
\end{equation}
Now to solve this, we start with a Gaussian ansatz of the following form for $W(X, p, t)\,.$
\begin{equation} \label{timwig}
    W(z, t) = \frac{1}{2\pi \sqrt{\det{\Sigma(t)}}} \exp{-\frac{(z-\bar{z}(t))^T \Sigma^{-1}(t) (z-\bar{z}(t))}{2}}\,,
\end{equation}
with $z = (X, p)\,.$ We also need the initial condition at $t=0\,.$ We start with an initial Gaussian state as evident from (\ref{densitynew1}).} Thereby we get
\begin{equation} \label{ininew}
    W(z,0) = \frac{1}{2\pi \sqrt{\det{\Sigma(0)}}} \exp{-\frac{z^T \Sigma^{-1}(0) z}{2}}\,.
\end{equation} 
 and 
\begin{equation}
    \Sigma(0) = 
    \begin{pmatrix}
        \sigma^2 & 0 \\
        0 & \frac{1}{4 \sigma^2}
    \end{pmatrix}.
\end{equation}}
\textcolor{black}{For our initial density matrix, the mean $\bar{z}(t=0)$ is zero as can be seen from (\ref{ininew}). So when solving for the time evolved mean position and momentum $\bar{X}(t)$ and $\bar{p}(t)$, which come from the well-known equations of a simple harmonic oscillator, we need to impose this initial condition. The solution takes the following form,  
\begin{equation}
    \bar{z}(t) = A \cos{\omega t} + B \sin{\omega t}\,.
\end{equation}
\textcolor{black}{Then the above mentioned initial condition necessitates that the coefficients $A, B = 0\,.$ Hence $\bar{z}(t) = \Vec{0}$. Thus, all that remains is solving for the covariance matrix $\Sigma(t)$ with zero means.}
\begin{align}
    \sigma_{XX} &= \langle X^2 \rangle\,,\quad 
    \sigma_{pp}= \langle p^2 \rangle, \quad 
    \sigma_{Xp} = \frac{1}{2} \langle X\,p + p\,X \rangle\,.
\end{align}
\textcolor{black}{In the Wigner representation, the expectation values are defined as, }
\begin{equation}
    \langle f(X, p) \rangle (t) = \int dX dp\, f(X, p)\, W(X, p, t)\,.
\end{equation}
Thus 
\begin{align}
    \frac{d}{dt}\sigma_{XX}(t) &= \int dX dp \, X^2\, \partial_t W(X, p, t)\,, \quad
    \frac{d}{dt}\sigma_{pp}(t) = \int dX dp\, p^2\, \partial_t W(X, p, t)\,,\nonumber\\
    \frac{d}{dt}\sigma_{Xp}(t) & = \frac{1}{2}\int dX dp\, (X\,p + p\,X)\, \partial_t W(X, p, t)\,.
\end{align}
Inserting this into (\ref{Wignerdiff}), we obtain the following differential equations.
\begin{align}\label{systemdiff}
    \dot{\sigma}_{XX}(t) &= 2 \sigma_{Xp}(t)\,,\quad 
    \dot{\sigma}_{pp}(t) = -2 \omega^2 \sigma_{Xp}(t) + 4 \gamma T\,,\quad
    \dot{\sigma}_{Xp}(t) = \sigma_{pp}(t) - \omega^2 \sigma_{XX}(t).
\end{align}
\textcolor{black}{Solving this system of differential equations (\ref{systemdiff}) we get (after imposing the fact that, $\sigma_{XX}(t=0)=\sigma^2\,, \sigma_{pp}(t=0)=\frac{1}{4\sigma^2}$ and $\sigma_{Xp}(t=0)=0$), }
\begin{align}
    \sigma_{XX}(t) &= \sigma^2 + \frac{2 \gamma T}{\omega^2} \left(t - \frac{\sin(2\omega t)}{2\omega} \right)\,,\nonumber\\
    \sigma_{pp}(t) &= \frac{1}{4 \sigma^2} + 2 \gamma T\left (t + \frac{\sin(2\omega t)}{2\omega} \right)\,,\nonumber\\ 
    \sigma_{Xp}(t) &=  \frac{\gamma T}{ \omega^2}\left(1 - \cos(2 \omega t )\right)\,.
\end{align}
\textcolor{black}{Subsequently, we get the time evolved Wigner function $W(z, t)$ from (\ref{timwig}) using the solutions of  $\sigma_{XX}(t)$, $\sigma_{Xp}(t)$ and $\sigma_{pp}(t)\,.$ \textcolor{black}{Then by performing an inverse Fourier transform, we return to the expression for the time evolved density matrix $\rho_s(x, x', t)$ with the help of (\ref{fourden}) in the original coordinate\,.} 
\textcolor{black}{\begin{equation}
    \rho_s(x, x', t) = \frac{1}{\sqrt{2 \pi \sigma_{XX}(t)}}\exp{-\alpha(t) \frac{(x + x')^2}{4} - \beta(t)(x-x')^2 + i \frac{\lambda(t)}{2}(x + x')(x - x') }\,,
\end{equation}
\textcolor{black}{with $$\alpha(t) = \frac{1}{2 \sigma_{XX}(t)}\,,\quad \beta(t) = \frac{1}{2}\left( \sigma_{pp}(t) - \frac{\sigma^2_{Xp}(t)}{\sigma_{XX}(t)}\right)\,,\quad \lambda(t) = \frac{\sigma_{Xp}(t)}{\sigma_{XX}(t)}\,.$$} 
Then we compute the following correlation function.}}
\begin{align}
    C(t) &= \frac{\Tr{\rho_s(t) \rho_s(0)}}{\Tr{\rho_s^2(0)}}= \int dx \int dx' \rho_s(x, x', t) \rho_s(x, x', 0)\,.
\end{align}
\textcolor{black}{Now we have to perform a simple Gaussian integral. Therefore the correlation function $C(t)$ becomes,}
\begin{equation}
    C(t) = \frac{1}{\sqrt{4 \alpha(t) \beta(t) - \lambda^2(t)}}\,.
\end{equation}
\textcolor{black}{Then, following the procedure outlined in Sec.~(\ref{KryComp}}), we can compute the first few Lanczos coefficients, and they are shown in Fig.~(\ref{fig:vnew5})\,.
\begin{figure}[H]
\centering
\includegraphics[width=.45\textwidth]{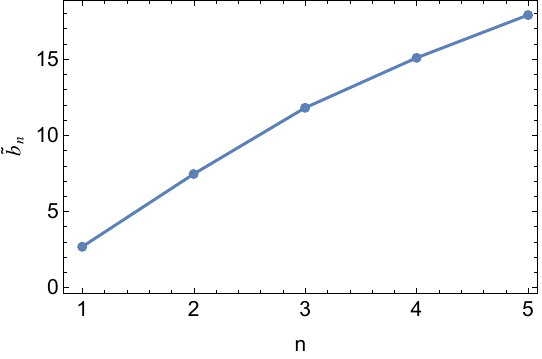}
\qquad
\includegraphics[width=.45\textwidth]{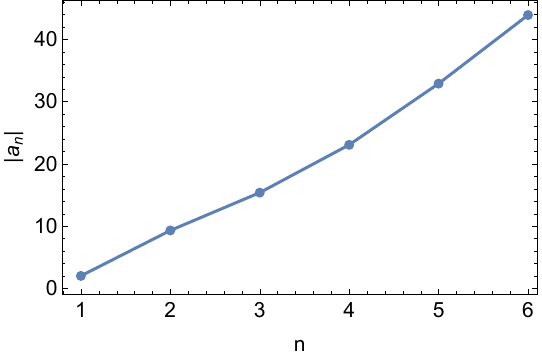}
\caption{Left: $b_n$ coefficients for $\omega = 1$, $\gamma = 1/10$, $\sigma = \sqrt{1/2}$ and $T = 300$. Right: absolute values of $|a_n|$  coefficients for the same parameters.}\label{fig:vnew5}
\end{figure}
The behavior of the Lanczos coefficients for the density matrix differs substantially from the Lanczos coefficients for the operators. The coefficients grow with $n$ while the $a_n$ coefficients are not complex, they are in fact real negative numbers (we have shown their magnitudes in the right panel of Fig.~(\ref{fig:vnew5})). Due to computational limitations, we are unable to obtain a large number of Lanczos coefficients, and consequently, the Krylov complexity can only be analyzed at early times, as shown in Fig.~(\ref{fig:12}). 
\begin{figure}[htb!]
\centering
\includegraphics[width=.45\textwidth]{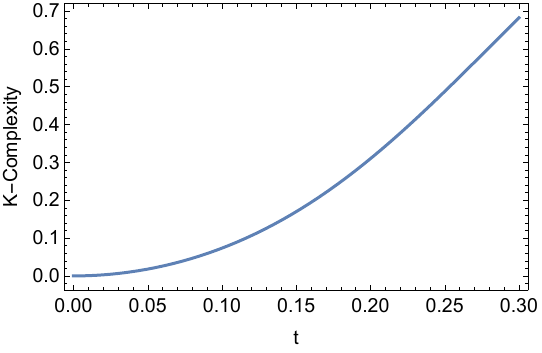}
\caption{Left: $\omega = 1$, $\gamma = 1/10$, $\sigma = \sqrt{1/2}$ and $T = 300$.}\label{fig:12}
\end{figure}
The onset of decoherence is generally an early-time feature in open quantum systems. However, despite the density matrix having a significantly larger number of Lanczos coefficients and a different Krylov complexity profile, we do not observe any clear early-time signature exclusively indicative of decoherence.}

\section{Discussion}
\label{Discussion}
In this work, we investigate Krylov complexity in open quantum systems, focusing on the damped harmonic oscillator and the Caldeira–Leggett model. We emphasize that the choice of relatively simple models is deliberate, providing the analytical tractability (especially for evaluating correlation functions) essential to dissect the roles of dissipation and decoherence via Krylov techniques. Although Krylov complexity and associated Lanczos coefficients have traditionally been leveraged to explore quantum chaos, this study re-purposes them to pinpoint these effects in open quantum systems. \textit{As far as we are aware, ours is an early effort to deploy Krylov methods for this purpose, shedding new light on the intricate interplay between dissipation and decoherence}. Our results show that Krylov complexity effectively captures dissipative and decoherence effects, revealing distinct behaviors in these systems. In the Caldeira–Leggett model, the late-time Krylov complexity exhibits saturation, indicative of a fully mixed state driven by dissipation and decoherence, but it does not clearly reflect the onset of decoherence. In contrast, for the damped harmonic oscillator, where decoherence is suppressed, Krylov complexity highlights pronounced dissipative effects, characterized by a sharp initial increase followed by oscillatory decay. These findings underscore the potential of Krylov complexity as a diagnostic tool for probing open quantum systems and open avenues for further exploration in quantum information theory and quantum Brownian motion.\\


Our analysis of the Caldeira-Leggett master equation reveals significant differences in the contributions of its individual terms, which account for dissipation and decoherence. As a preliminary model, we employed the damped harmonic oscillator master equation to isolate the features of dissipation in Krylov complexity. In this case, Krylov complexity exhibits a rapid initial rise followed by an oscillatory decay, reflecting the dissipative dynamics. In the C-L model, which contains distinct terms for dissipation and decoherence, we found that with appropriate choices for the operator and in the high-temperature limit, these terms can be independently suppressed. When the decoherence term is turned off, the results mirror those of the damped harmonic oscillator, showing a sharp increase and oscillatory decay driven by dissipation. Conversely, when the dissipative term is suppressed, the dynamics align with the coherent behavior described in Appendix~(\ref{AppA}). In this scenario, the decoherence term in the C-L master equation primarily drives operator diffusion rather than dissipation, leading to oscillatory behavior with periodic ``bottlenecks'' caused by a second oscillatory frequency introduced by decoherence. Increasing the temperature in the decoherence term amplifies the frequency of these oscillations, manifesting as ``noise'' in the Krylov complexity, characterized by rapid oscillations and a slow decay due to the interplay of the second frequency mode.  \textcolor{black}{Note that the C-L master equation considers a natural continuation with a more complex Lindblad operator. The Lindblad operator for the C-L model is a linear combination of the creation and annihilation operators used in the preliminary damped oscillator model \footnote{\textcolor{black}{In fact other combinations of creation and annihilation operators  of harmonic oscillator as Lindblad operators will not necessarily generate additional interesting results beyond what is reported in the manuscript due to the underlying algebra of simple harmonic oscillator.}}. While this is not necessarily the same as considering multiple Lindblad operators, the choice of model and hence the Lindblad operators is determined by the model itself, allowing us to focus on the evolution of Gaussian states and making the computation of moments, and subsequently the Krylov complexity, tractable. Although it will be interesting to generalize our studies for models that will provide us with more Lindblad operators (and possibly taking the evolved state beyond Gaussian), some of the conclusions, e.g. suitability of Krylov basis to detect decoherence in Sec.~(\ref{newsec22}), reached in this paper, we believe nevertheless will remain as it is as they come from generic arguments about the choice of the basis and independent of the choice of jump operators.} \\

\textcolor{black}{Also, we have investigated the early time growth of the Krylov complexity for density matrices keeping only the decoherence term in the C-L master equation, leaving the analysis for full master equation for a future study (refer to Appendix~(\ref{AppCnew}) for a sketch of the computation of the moments for this case). We can only manage to compute a few Lanczos coefficients, but unlike the operator case, they do not truncate quickly. While the dimension of the Krylov space is much larger (than the case of the operator) and the behaviour of the Lanczos coefficients is different for the density matrix, there is still no early time signature exclusively indicative of decoherence in the Krylov complexity. We hope to do a more thorough analysis of the Krylov complexity for the density matrix by keeping both the dissipative and the decoherence term in C-L master equation in the near future.} \\

We compare our Krylov complexity results for the Caldeira–Leggett model with those of \cite{Bhattacharyya_2023}, where circuit complexity based on Nielsen’s geometric approach was used to compute the complexity of purification for the reduced density matrix in a system coupled to a free scalar field bath. Both Krylov and circuit complexity exhibit saturation, signaling a fully mixed state due to decoherence. In \cite{Bhattacharyya_2023}, it was noted that circuit complexity saturates on a timescale similar to that of linear entropy and entanglement negativity. By contrast, our Krylov complexity results saturate at a later stage, once decoherence becomes dominant, compared with the onset indicated by the decoherence function. This suggests that circuit complexity is more sensitive to detecting the onset of mixedness, while Krylov complexity highlights operator growth as decoherence progresses.\\

This relative insensitivity of Krylov complexity to the early stages of decoherence can be traced to the choice of basis. In particular, the behavior of the off-diagonal terms in the density matrix strongly depends on the so-called preferred basis \cite{Schlosshauer_2019,BRE02}. The Krylov basis required for the Krylov complexity does not satisfy the conditions of the preferred basis, which may explain why our results for the Caldeira–Leggett master equation do not clearly reproduce the decoherence signatures observed in prior studies. Future work could explore how operator growth defined in the preferred basis might better capture decoherence signals, thereby enhancing the utility of operator complexity in open quantum systems.\\

In parallel with this line of inquiry, related extensions are also underway. For instance, in \cite{Bhattacharyya_2024}, the Complexity of Purification was analyzed for the Caldeira–Leggett model in curved space, revealing additional distinctive features. It would be interesting to examine whether similar behavior also appears in the context of Krylov operator complexity, and we are currently investigating this question. Another promising direction is to study recoherence phenomena in these curved-space models \cite{Bhattacharyya:2025cxv}.
\newpage

\section*{Acknowledgments}
The authors would like to thank HJR Van Zyl and Arpon Paul for discussions. AB would like to thank the speakers and participants of the BIRS-CMI workshop (25w5386) ``Quantum Gravity and Information Theory: Modern Developments" (2025) for stimulating discussion. SSH is supported in part by the National Institute for Theoretical and Computational Sciences of South Africa (NITheCS). AB is supported by the Core Research Grant (CRG/2023/ 001120) by the Department of Science and Technology (DST) and Anusandhan National Research Foundation (ANRF), India. MSG is supported by the National Research Foundation postgraduate bursary. AB also acknowledges the associateship program of the Indian Academy of Sciences (IASc), Bengaluru, and acknowledges support from the Indian Institute of Technology Gandhinagar and a generous donor through the Singheswari and Ram Krishna Jha Chair.

\appendix


\section{Krylov Complexity for the Harmonic Oscillator}\label{AppA}
\begin{figure}[H]
\centering
\includegraphics[width=.45\textwidth]{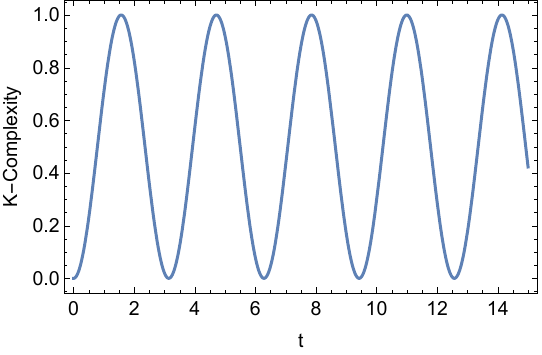}
\qquad
\caption{Time evolution of Krylov complexity for Harmonic oscillator with $\omega = 1$ and $m = 1$. \label{fig:A2}}
\end{figure}

Recall that from (\ref{eq:operator_evolution}) the operator evolution is given as:
\begin{align}\label{11}
    O(t) &= O + (it)[H, O] + \frac{(it)^2}{2!}[H,[H, O]] + \frac{(it)^3}{3!}[H,[H,[H, O]]] + \ldots 
\end{align}
We begin with the operator $O = \alpha(a^{\dag}+a)$ and Hamiltonian $H = \omega(a^\dag a +\frac{1}{2})$. Then the first few commutators in (\ref{11}) is calculated as:
\begin{align}\label{2.10}
    [H,O] &= \omega \alpha(a^{\dag} - a)\,, 
    [H,[H,O]] = \omega^2 \alpha(a^{\dag} + a)\,,
    [H,[H,[H,O]]] = \omega^3 \alpha (a^{\dag} - a)\,. 
\end{align}
From this we can see that there is clearly a pattern. Then inserting this back into (\ref{11}) we get:
\begin{align}\label{12}
    O(t) &=  \alpha(a^{\dag}+a) + (it \omega) \alpha(a^{\dag} - a) + \frac{(it\omega)^2}{2!}\alpha(a^{\dag}+a) + \frac{(it\omega)^3}{3!}\alpha(a^{\dag}-a) + \ldots\,, \nonumber\\
    &= \left[\alpha(a^{\dag}+a)\left(1+\frac{(it\omega)^2}{2!} + \ldots \right)\right] + \left[\alpha(a^{\dag}-a)\left(it\omega +\frac{(it\omega)^3}{2!}+ \ldots \right)\right] \,, \nonumber\\
    &= \alpha\cos(\omega t)(a^{\dag}+a) + i \alpha \sin( \omega t)(a^{\dag}-a) 
    = \alpha(e^{i \omega t} a^{\dag} + e^{-i \omega t} a)\,.
\end{align}
Running this through the procedure of the moments method mentioned in Sec.~(\ref{KryComp}), as well as the Lanczos algorithm, we find the following result for the Krylov complexity, which we show in Fig.~(\ref{fig:A2}).

\section{Correlation functions and Time evolved operators}\label{AppB}

Here we show the general procedure followed to compute the time-evolved operators for the Caldeira-Leggett master equation in Sec.~(\ref{CLmod}). We only show the procedure followed in the first case, since for the rest of the operators, only a slight change is needed. We begin with the full form of the Lindblad master equation as given in (\ref{LindbladCL}), with the initial operator being $O(0) = p^2$, this is because it is the simplest form of operator that allows us to obtain contributions from both the dissipative and decoherence terms. 
The Lindblad jump operator $L$ is given in (\ref{lind}).
The general procedure we will follow here is similar to Sec.~(\ref{DampedOsc}). The evolution of such operators when written in terms of creation and annihilation operators can, in general, be written in terms of some time-dependent functions $f_i(t) \in \mathbb{C}$. As such, initially, we will leave the function, as well as the time dependence, implicit, focusing only on the operator algebra, and reintroduce the time dependence later. We begin by considering the individual contributions from each term of (\ref{LindbladCL}). From the first term, for the coherent dynamics,  we have:

\begin{align}
    i[H_s, p^2] &= i\left[\left(\frac{p^2}{2} + \frac{\omega^2 x^2}{2} \right),p^2 \right]=
     -\omega^2 xp - \omega^2 px\,.
\end{align}

The second term of (\ref{LindbladCL}), which behaves like a Hamiltonian contribution and comes from the dissipator, yields the following:

\begin{align}
    i\frac{\gamma}{2} [x\, p + p\,x, p^2] &= i \frac{\gamma}{2} [x\, p , p^2] + i \frac{\gamma}{2} [p\, x , p^2]
   = -  2 \gamma\, p^2\,.
\end{align} 

Now we tackle the contributions which come from the dissipator in the Lindblad master equation, the third term of the master equation (\ref{LindbladCL}) gives us:

\begin{align}
    L^\dag p^2 L &= (A x- i B p)(p^2)(Ax + i B p)= A^2 x\, p^2 x + i A B x\, p^3 + B^2 p^4 - i A B p^3\, x \,.
\end{align} 

From the fourth term of (\ref{LindbladCL}), we get:

\begin{align}
    - \frac{1}{2} L^\dag  L p^2&= (A x- i B p)(Ax + i B p) (p^2)= - \frac{1}{2} (A^2 x^2\, p^2 + i A B x\, p^3 - i A B p x\, p^2 + B^2 p^4)
\end{align} 

and finally, from the fifth term of (\ref{LindbladCL}) we get: 

\begin{align}
    - \frac{1}{2}p^2 L^\dag  L &= (p^2)(A x- i B p)(A\,x + i B\,p)= - \frac{1}{2} (A^2 p^2\, x^2 + i A B p^2\, x\, p - i A B p^3\, x + B^2 p^4)\,.
\end{align} 
Combining all of them, we get,

\begin{align}
    S &= \gamma\,( A^2 x\, p^2\, x - \frac{1}{2} A^2 x^2\, p^2 - \frac{1}{2} A^2 p^2\, x^2) + \gamma\,( \frac{i}{2} A B x\, p^3 - \frac{i}{2} A B p^3\, x) + \gamma\,( \frac{i}{2} A B p\, x\, p^2 - \frac{i}{2} A B p^2\, x\, p)\,,\nonumber \\
    &= \gamma\,( A^2 x\, p^2\, x - \frac{1}{2} A^2 x^2\, p^2 - \frac{1}{2} A^2 p^2\, x^2) + \gamma\, \frac{i A B}{2} [x, p^3] + \gamma\, \frac{i A B}{2} [p\,x\, p ,p]
    = \gamma( A^2 - 2ABp^2)\,.
\end{align}

This leaves us with the following:

\begin{align}
  \frac{d}{dt} O &\propto  -\omega^2 xp - \omega^2 px -  2 \gamma p^2 + \gamma( A^2 - 2ABp^2)\,, \nonumber\\
  &\propto -\omega^2 xp - \omega^2 px + (-  2 \gamma - \gamma 2AB)p^2 + \gamma A^2 (a a^\dag - a^\dag a)\,. \label{neweq}
\end{align}

This is the equation (\ref{maineq}) that we have used in the main text. Now to solve for the $O(t),$ we first recall that $a(t) = f^*(t) a$ and similarly $a^{\dag}(t) = f(t) a^{\dag}$ as discussed in Sec.~(\ref{DampedOsc}). We then rewrite every term of (\ref{neweq}) in terms of $a(t)$ and $a^{\dag}(t)$ in the following way,



\begin{align}
   (- \omega^2 ) x p &= \underbrace{(- \frac{i}{2} \omega^2 )}_M (a^\dag(t) a(t) - a^\dag(t) a^\dag(t) + a(t)a(t) - a(t) a(t)^\dag)\,,\nonumber \\ 
   (- \omega^2 ) p\,x &= \underbrace{(- \frac{i}{2} \omega^2 )}_N (a(t) a^\dag(t)  - a^\dag(t) a^\dag(t) + a(t)a(t) - a^\dag(t) a(t))\,,\nonumber \\ 
   (-2 \gamma - 2 \gamma AB) p^2 &= \underbrace{\frac{\omega}{2} (2 \gamma + 2 \gamma AB)}_O(a(t)a(t) - a(t) a^\dag(t) - a^\dag(t) a(t) + a^\dag(t) a^\dag(t))\,.
\end{align}

We now have the following form of the differential equation. 

\begin{align}
  \frac{d}{dt} O &=  -(- M + N + O) f_1(t) a^\dag a + (-M - N -O) f_2(t)a^\dag a^\dag \nonumber \\ &+ (M +N + O) f_3(t) aa - (M - N + O) f_4(t) a a^\dag \,.
\end{align}

The functions $f_1, f_2, f_3, f_4$ can be read off from (\ref{CreAnn}).  Finally, we get an explicit form of the time-evolved operator as:

\begin{align}
  O(t) &= -\frac{\omega}{2}\Big(\exp\Big[( 2i\omega   - 2 \gamma -2\gamma AB)t\Big]aa - \exp\Big[(  \gamma A^2- 2 \gamma  -2 \gamma AB)t\Big]a a^\dag \nonumber \\
  &- \exp\Big[( - \gamma A^2- 2 \gamma  -2 \gamma AB)t\Big] a^\dag a \nonumber + \exp\Big[(- 2i\omega -2 \gamma -2\gamma AB)t\Big] a^\dag a^\dag\Big)\,.
\end{align}

Equipped with this, we can therefore compute the correlator: 

\begin{align}
    \langle O(t) O \rangle_{\beta} &= \frac{\mathrm{Tr}\{\rho_\beta O(t) O \}}{\mathrm{Tr}\{\rho_{\beta} O^2\}}\,, \nonumber \\
    &= \frac{1}{\mathrm{Tr}\{\rho_{\beta} O^2\}}\sum_n  e^{-\beta(n+1/2)\omega} \langle n | \frac{\omega^2}{4}(e^{( 2i\omega   - 2 \gamma -2\gamma AB)t}aa
    - e^{(  \gamma A^2- 2 \gamma  -2 \gamma AB)t}a a^\dag \nonumber \\
    &- e^{( - \gamma A^2- 2 \gamma  -2 \gamma AB)t} a^\dag a 
    + e^{(- 2i\omega -2 \gamma -2\gamma AB)t} a^\dag a^\dag)(  a^{\dag}a^{\dag} - a^{\dag}a -  a a^{\dag} + aa) | n \rangle\,,\nonumber \\
    &= \frac{1}{\mathrm{Tr}\{\rho_{\beta} O^2\}} \frac{\omega^2}{4} \sum_n  e^{-\beta(n+1/2)\omega} ( e^{( 2i\omega   - 2 \gamma -2\gamma AB)t} (n^2+3n+2)\nonumber \\ &+ e^{(  \gamma A^2- 2 \gamma  -2 \gamma AB)t} (2n^2+3n+1) + e^{( - \gamma A^2- 2 \gamma  -2 \gamma AB)t} (2n^2+n) \nonumber\\ &+ e^{(- 2i\omega -2 \gamma -2\gamma AB)t} (n^2-n))\,.
\end{align}

Here, the choice of normalization is the common choice for normalization to ensure that at $C(0) = 1\,.$ Now we use the following to perform the sum over $n$, 

\begin{align}
\begin{split}
   & \sum_n \langle n | a^{\dag}a^{\dag}aa | n \rangle = n^2-n\,,\quad  \sum_n \langle n | aa a^{\dag}a^{\dag} | n \rangle = n^2+3n+2\,, \nonumber \\& \sum_n \langle n | a a^{\dag} a a^{\dag} | n \rangle = n^2+2n+1\,,\quad \sum_n \langle n | a a^{\dag} a^{\dag} a | n \rangle = n^2+n\,, \\& \sum_n \langle n | a^{\dag} a a^{\dag} a | n \rangle = n^2\,,\quad \sum_n \langle n | a^{\dag} a a a^{\dag}| n \rangle = n^2+n\,.
   \end{split}
\end{align}


Finally, one can use the expressions for $C_1, C_2, C_3$ as mentioned in (\ref{eq2}) of Sec.~(\ref{CLmod}) to bring it into a more compact form. 
The procedure outlined above was used in computing the operators $O(t)$ and correlators for the entirety of Sec.~(\ref{CLmod}). Then we use (\ref{new11}) and (\ref{new111})) to compute the Lanczos coefficients. 

\textcolor{black}{\section{Parameter dependence of the Lanczos coefficients in the C-L model} \label{appc}
Here, we will check the parameter dependence of the Lanczos coefficients as well as the dimension of the Krylov space. We observe that changing the parameters $\omega$ and $\gamma$ does not change the Krylov dimension (Lanczos coefficients still terminate after $n=5$ as shown in Fig.~(\ref{fig:LanczosCoeff}) and (\ref{fig:LanczosCoeff1}) of the main text), but it results in a scaling of the coefficients as can be seen from the Fig.~(\ref{fig:vnew4}) and Fig.~(\ref{fig:vnew3}). $\tilde{b}_n's$ remain real and $a_n's$ are complex like before, and we have plotted their imaginary and real parts in  Fig.~(\ref{fig:vnew3})\,.
\begin{figure}[htb!]
\centering
\includegraphics[width=.44\textwidth]{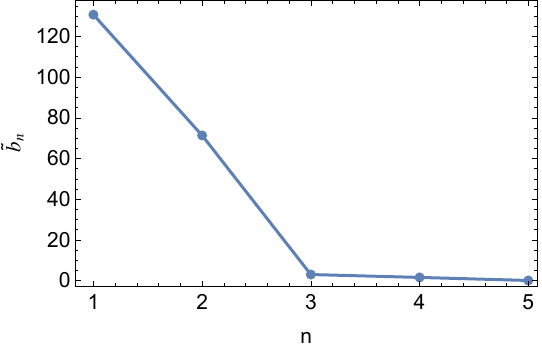}
\qquad
\includegraphics[width=.44\textwidth]{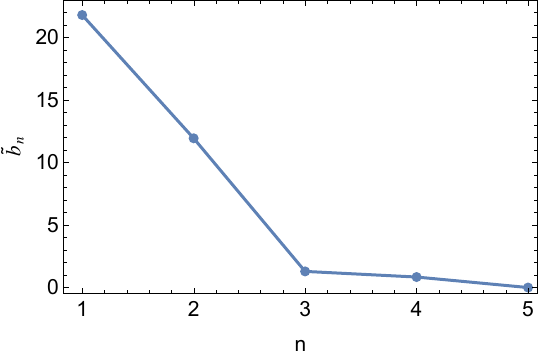}
\caption{Left: $\tilde{b}_n$ coefficients for $\omega = 1$ and $\gamma = 1/5$. Right: $\tilde{b}_n$ coefficients for $\omega = 1/2$ and $\gamma = 1/15$.}\label{fig:vnew4}
\end{figure}
\begin{figure}[htb!]
\centering
\includegraphics[width=.44\textwidth]{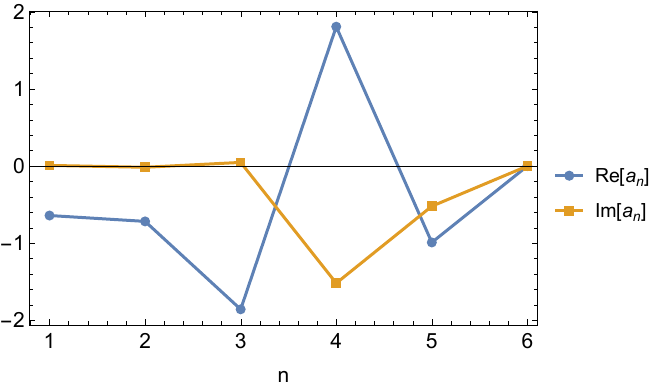}
\qquad
\includegraphics[width=.44\textwidth]{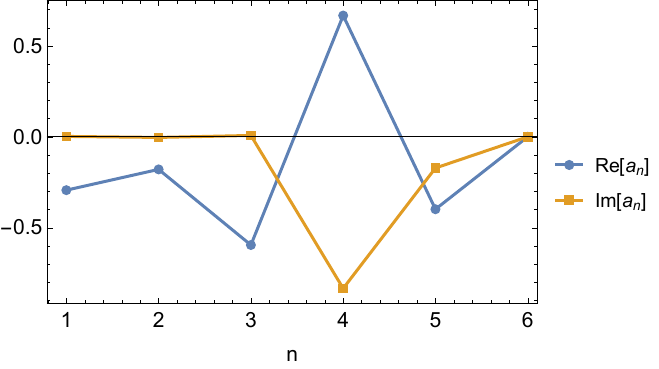}
\caption{Left: $a_n$ coefficients for $\omega = 1$ and $\gamma = 1/5$. Right: $a_n$ coefficients for $\omega = 1/2$ and $\gamma = 1/15$.}\label{fig:vnew3}
\end{figure}}
\textcolor{black}{\section{Krylov elements for the C-L model}\label{appD}
\textcolor{black}{Here we show the plots of the Krylov elements (probability amplitudes $\phi_n$) for the C-L model discussed in Sec.~(\ref{secnew1}). We clearly see that, as also discussed in \cite{Bhattacharya_2023}, for open systems the $\sum_n |\phi_n|^2 \neq 1, \forall t$ reflects the loss of information from the system to the environment, as a result of the non-unitary dynamics of the open system.}
\begin{figure}[H]
\centering
\includegraphics[width=.40\textwidth]{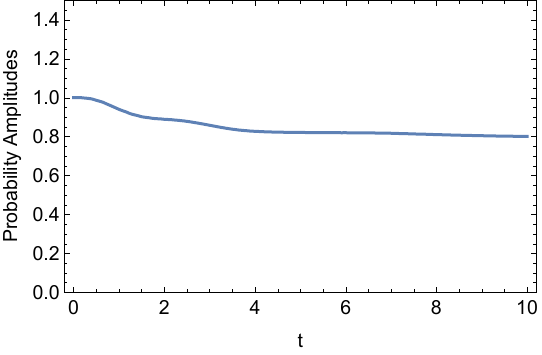}
\includegraphics[width=.40\textwidth]{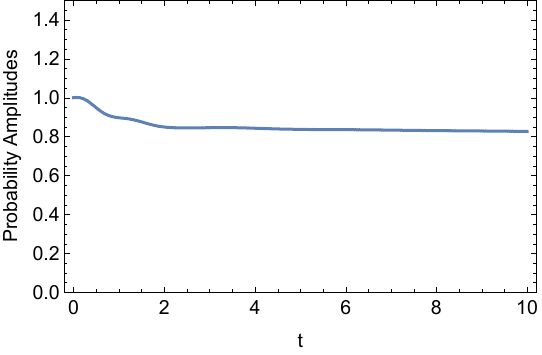}
\qquad
\caption{Probability amplitudes for $\omega = 1$ and $\gamma = 1/15\,,\gamma = 1/10\,.$}\label{fig:vnew2}
\end{figure}
For the sake of completeness, we have also shown a few of the Krylov elements $\phi_n$ as a function of time in Fig.~(\ref{fig:vnew32}). We have used (\ref{eq:schrodinger_like}) to compute them and then used them in (\ref{comnew}) to get the Krylov complexity, which was discussed in Sec.~(\ref{secnew1}).
\begin{figure}[H]
\centering
\includegraphics[width=.45\textwidth]{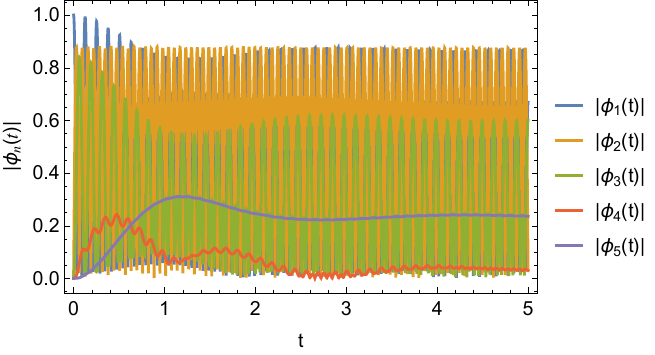}
\includegraphics[width=.45\textwidth]{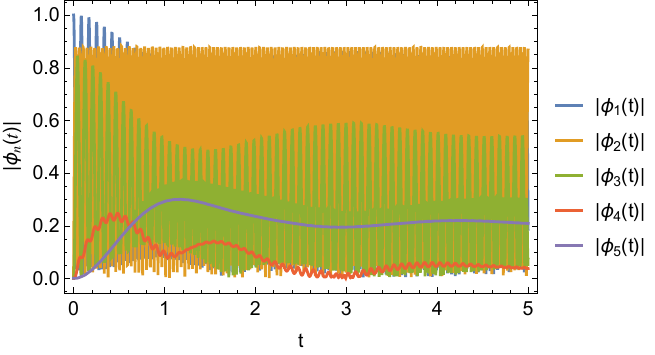}
\qquad
\caption{$|\phi_n(t)|$ for $\omega = 1$ and $\gamma = 1/15\,,\gamma = 1/10\,.$}\label{fig:vnew32}
\end{figure}}



\textcolor{black}{\section{Computation of the moments for the Density Matrix evolved by the full C-L master equation}\label{AppCnew}
Here we consider the full master equation as opposed to only the double commutator decoherence term in Sec.~(\ref{4.4.4}). The C-L master equation (with $\hbar = 1$) is
\begin{equation}
\frac{d\rho_s}{dt}
= -i[H_s, \rho_s] - i\gamma[x, \{p, \rho_s\}] - D[x, [x, \rho_s]],
\end{equation}
where $D = 2 m \gamma k_B T$. In the position representation $\rho_s(x, x', t)$, using $p = -i \partial_x$, we obtain:
\begin{align}
-i[H_s, \rho_s] &= -i \left[
- \frac{1}{2m}(\partial_x^2 - \partial_{x'}^2)
+ \frac{1}{2} m \omega^2 (x^2 - x'^2)
\right] \rho_s\,,\nonumber \\
-i\gamma[x, \{p, \rho_s\}] &= -\gamma (x - x')(\partial_x - \partial_{x'}) \rho_s\,, \nonumber\\
-D[x, [x, \rho_s]] &= -D (x - x')^2 \rho_s\,.
\end{align}
Introduce sum and difference coordinates:
\begin{equation}
r = x - x', \qquad s = \frac{x + x'}{2}\,.
\end{equation}
Then
\begin{align}
\partial_x &= \partial_r + \frac{1}{2}\partial_s\,.\quad
\partial_{x'} = -\partial_r + \frac{1}{2}\partial_s\,.
\end{align}
Then the master equation becomes,
\begin{equation}\label{EQ1}
\frac{\partial \rho_s}{\partial t}
= \frac{i}{m} \partial_r \partial_s \rho_s
- i m \omega^2 r s \rho_s
- 2 \gamma r \partial_r \rho_s
- D r^2 \rho_s\,.
\end{equation}
The initial state is a Gaussian wave packet:
\begin{equation}
\rho_s(x, x', 0)
= \frac{1}{\sqrt{2\pi \sigma^2}}
\exp\left[-\frac{x^2 + x'^2}{4\sigma^2}\right].
\end{equation}
In $(r,s)$ coordinates it takes the following form,
\begin{equation}
\rho_s(r,s,0)
= \frac{1}{\sqrt{2\pi \sigma^2}}
\exp\left[-\frac{s^2}{2\sigma^2} - \frac{r^2}{8\sigma^2}\right]
\end{equation}
with,
\begin{equation}
\alpha_0 = \frac{1}{8\sigma^2}, \quad
\beta_0 = \frac{1}{2\sigma^2}, \quad
\gamma_0 = 0, \quad
\delta_0 = -\frac{1}{2}\ln(2\pi \sigma^2).
\end{equation}
To find the time-evolved density matrix, we again use a Gaussian ansatz as the Lindbladian is at most quadratic. 
\begin{equation}
\rho_s(r,s,t)
= \exp\left[
-\alpha(t) r^2 - \beta(t) s^2 - \gamma(t) r s + \delta(t)
\right]\,. \label{appnew}
\end{equation}
Imposing the trace preservation condition, we get, 
\begin{equation}
\int \rho_s(r=0, s, t)\, ds = 1
\quad \Rightarrow \quad e^{\delta} = \sqrt{\frac{\beta}{\pi}}\,.
\end{equation}
Substituting into Eq.(\ref{EQ1}) yields the following, 
\begin{align}
\dot{\alpha} &= -\frac{i}{m}(2\alpha\gamma) - 4\gamma \alpha + D\,,
\dot{\beta}= -\frac{i\,2\beta \gamma}{m}\,, \nonumber\\
\dot{\gamma} &= -\frac{i}{m}(4\alpha\beta + \gamma^2)
+ i m \omega^2 + 2\gamma^2\,, \quad 
\dot{\delta} = -\frac{i\,\beta \gamma}{m}\,. 
\end{align}
Then we define the correlation matrix as,
\begin{equation}
C(t) = \frac{\mathrm{Tr}[\rho_s(t)\rho_s(0)]}{\mathrm{Tr}[\rho_s(0)^2]}\,.
\end{equation}
Since $\mathrm{Tr}[\rho_s(0)^2]=1$,
\begin{equation}
C(t) = \iint \rho_s(r,s,t)\rho_s(r,s,0)\, dr\, ds\,.
\end{equation}
This takes the following form after using (\ref{appnew}),
\begin{equation}
C(t)
= \sqrt{\beta(t)\beta_0}
\left[
(\alpha(t)+\alpha_0)(\beta(t)+\beta_0) - \frac{\gamma(t)^2}{4}
\right]^{-1/2}.
\end{equation}
Now we know that, at $t=0$, $\gamma(0)=0$ and, 
\begin{align}
\dot{\alpha}(0) &= D - 4\gamma \alpha_0\,,\quad 
\dot{\beta}(0)= 0\,,\nonumber\\
\dot{\gamma}(0)&= -i\left(\frac{4\alpha_0 \beta_0}{m} - m\omega^2\right)\,,\quad 
\dot{\delta}(0)= 0\,.
\end{align}
We also quote the second derivatives (at t=0), as they will be required for the moment computation.
\begin{align}
\ddot{\alpha}(0) &= -\frac{8\alpha_0^2 \beta_0}{m^2}
+ 2\alpha_0 \omega^2 - 4\gamma \dot{\alpha}(0)\,, \nonumber\\
\ddot{\beta}(0) &= -\frac{8\alpha_0 \beta_0^2}{m^2}
+ 2\beta_0 \omega^2\,,\ddot{\gamma}(0) = -\frac{4i}{m} \dot{\alpha}(0)\beta_0\,.
\end{align}
Then following Sec.~(\ref{KryComp}) we get the moment as follows, 
\begin{equation}
\mu_n = \left.\frac{d^n C}{dt^n}\right|_{t=0}.
\end{equation}
So it is evident that we will only require the time derivatives at $t=0$, and they are already computed above. Below we quote the first few moments.\\\\ 
{\bf First moment:}
\begin{equation}
\mu_1 = -\frac{\dot{\alpha}(0)}{4\alpha_0}
= -\frac{D}{4\alpha_0}
= -2 D \sigma^2\,.
\end{equation}
{\bf Second moment:}
\begin{equation}
\frac{\ddot{C}}{C}\bigg|_{t=0}
=
\left(\frac{\dot{C}}{C}\right)'\Bigg|_{t=0}
+
\left(\frac{\dot{C}}{C}\right)\Bigg|_{t=0}^2.
\end{equation}
Let's define the following, 
\begin{equation}
F = (\alpha + \alpha_0)(\beta + \beta_0) - \frac{\gamma^2}{4}\,.\nonumber
\end{equation}
Then we have the second moment as, 
\begin{align}
\mu_2 =
\frac{1}{2}\frac{\ddot{\beta}(0)}{\beta_0}
- \frac{1}{2}
\left[
\frac{\ddot{F}(0)}{F_0}
- \left(\frac{\dot{F}(0)}{F_0}\right)^2
\right]
+ \mu_1^2\,,
\end{align}
with
\begin{align}
F_0= 4\alpha_0 \beta_0\,,
\dot{F}(0)= 2\beta_0 \dot{\alpha}(0)\,,
\ddot{F}(0)= 2\beta_0 \ddot{\alpha}(0)
+ 2\alpha_0 \ddot{\beta}(0)
- \frac{1}{2}\dot{\gamma}(0)^2\,.
\end{align}
Higher moments can be similarly computed and then the Lanczos coefficients from them numerically using (\ref{new11}) and (\ref{new111}).}

\bibliographystyle{JHEP}
\bibliography{Biblio.bib}

\end{document}